\documentclass{aastex}
\usepackage{emulateapj5}

\usepackage{epsfig}

\def \kid{$\chi^2$}
\def \deltakid{$\Delta\chi^2$}

\def \lya{Ly-$\alpha$}
\def \lyb{Ly-$\beta$}
\def \lyg{Ly-$\gamma$}
\def \lyd{Ly-$\delta$}

\def \cmmd{cm$^{-2}$}
\def \kms{${\rm km\,s}^{-1}$}

\def \dshism{(D/H)$_{\rm{ISM}}$}
\def \dshpe{(D/H)$_{\rm{PE}}$}

\def \dsolb{(D/O)$_{\rm{LB}}$}
\def \ow{Owens.f}
\def \fuse{{\it FUSE}}
\def \2s{$2\,\sigma$}
\def \1s{$1\,\sigma$}
\def \lsf{line spread function}
\def \hd9{HD\,90087}
\def \feige{Feige\,110}

\shorttitle{{\it FUSE} determination of a low deuterium abundance along 
an extended sight line in the Galactic disk}
\shortauthors{H\'ebrard et al.}

\begin{document}

\title{{\it FUSE} determination of a low deuterium abundance along an 
extended sight line in the Galactic disk}

\author{
G.~H\'ebrard\altaffilmark{1}, T.~M.~Tripp\altaffilmark{2},
P.~Chayer\altaffilmark{3,4}, S.~D.~Friedman\altaffilmark{5},
J.~Dupuis\altaffilmark{3},
P.~Sonnentrucker\altaffilmark{3}, 
G.~M.~Williger\altaffilmark{3},
H.~W.~Moos\altaffilmark{3}}

\altaffiltext{1}{Institut d'Astrophysique de Paris,
UMR7095 CNRS, Universit\'e Pierre \& Marie Curie, 98$^{bis}$ boulevard
Arago, F-75014 Paris, France; \\hebrard@iap.fr.}

\altaffiltext{2}{Department of Astronomy, University of Massachusetts, 
Amherst, MA 01003, USA.}

\altaffiltext{3}{Department of Physics \& Astronomy, 
The Johns Hopkins University, Baltimore, MD 21218, USA.}

\altaffiltext{4}{Department of Physics \& Astronomy, 
University of Victoria, P.O. Box 3055, Victoria, BC V8W 3P6, Canada.}

\altaffiltext{5}{Space Telescope Science Institute, 3700 San Martin Drive, 
Baltimore, MD 21218, USA.}

\begin{abstract} 
We present a study of the deuterium abundance along the extended sight
line toward \hd9\ with the {\it Far Ultraviolet Spectroscopic
Explorer} ({\it FUSE}). \hd9\ is a O9.5III star located in the
Galactic disk at a distance of $\sim2.7$~kpc away from the Sun.  Both
in terms of distance and column densities, \hd9\ has the longest and
densest sight line observed in the Galactic disk for which a deuterium
abundance has been measured from ultraviolet absorption lines so far.
Because many interstellar clouds are probed along this sight line,
possible variations in the properties of individual clouds should be
averaged out.  This would yield a deuterium abundance which is
characteristic of the interstellar medium on scales larger than the
Local Bubble.  The \fuse\ spectra of \hd9\ show numerous blended
interstellar and stellar features.  We have measured interstellar
column densities of neutral atoms, ions, and molecules by
simultaneously fitting the interstellar absorption lines detected in
the different \fuse\ channels.  As far as possible, saturated lines
were excluded from the fits in order to minimize possible systematic
errors. {\it IUE} ({\it International Ultraviolet Explorer}) archival
data are also used to measure neutral hydrogen. We report
D/O~$=(1.7\pm0.7)\times10^{-2}$ and D/H~$=(9.8\pm3.8)\times10^{-6}$
(\2s).  Our new results confirm that the gas-phase deuterium abundance
in the distant interstellar medium is significantly lower than the one
measured within the Local Bubble.  We supplement our study with a
revision of the oxygen abundance toward \feige, a moderately distant
($\sim200$~pc) sdOB star, located $\sim150$~pc below the Galactic
plane. Excluding saturated lines from the fits of the \fuse\ spectra
is critical; this led us to derive an \ion{O}{1} column density about
two times larger than the one previously reported for \feige. The
corresponding updated D/O ratio on this sight line is D/O~$=(2.6
\pm1.0) \times 10^{-2}$ (\2s), which is lower than the one measured
within the Local Bubble.  The dataset available now outside the Local
Bubble, which is based primarily on \fuse\ measurements, shows a
contrast between the constancy of D/O and the variability of D/H. As
oxygen is considered to be a good proxy for hydrogen within the
interstellar medium, this discrepancy is puzzling.
\end{abstract}


\section{Introduction}
\label{Introduction}

Deuterium, which is produced during primordial nucleosynthesis and
then destroyed by astration, is a key element in cosmology.  Whereas
measurements in low-metallicity QSO absorption systems probe D/H at
look-back times of $\sim13-14$~Gyrs, the present epoch deuterium
abundance \dshpe\ can be measured in the interstellar medium (ISM).

The \fuse\ ({\it Far Ultraviolet Spectroscopic Explorer}) mission has
brought significant progress on the \dshism\ measurement. D/H likely
has a single value in the Local Bubble\footnote{The Local Bubble is a
$\sim100$~pc-size low-density cavity in which the Solar System is
embedded (see, i.e., Sfeir et al.~\citealp{sfeir99})}, in the range
$(1.3-1.5) \times 10^{-5}$ (Moos et al.~\citealp{moos02}; H\'ebrard \&
Moos~\citealp{hebrard03}). However, it now appears that this local
abundance should not be considered as a canonical value characteristic
of the Milky Way at the present epoch, as it was usually believed
before \fuse\ studies.  Indeed, numerous distant sight lines,
i.e. sight lines probing the interstellar medium beyond the Local
Bubble, have shown deuterium abundances in disagreement with the Local
Bubble one (see, e.g, Laurent et al.~\citealp{laurent79}, Jenkins et
al.~\citealp{jenkins99}, Lemoine et al.~\citealp{lemoine99}, Sonneborn
et al.~\citealp{sonneborn00}, Hoopes et al.~\citealp{hoopes03}).

H\'ebrard \& Moos~(\citealp{hebrard03}) reported a trend in the
deuterium abundance based on D/O, D/N, and previously reported D/H
measurements: the deuterium abundance is lower than in the Local
Bubble for the most distant sight lines exhibiting the highest
hydrogen column densities. More recent work by Wood et
al.~(\citealp{wood04}) confirms this low value.  These results suggest
that the deuterium abundance might be locally abnormally higher than
the present epoch value. While H\'ebrard \& Moos~(\citealp{hebrard03})
suggested that the present epoch ratio \dshpe\ would be significantly
lower than $1\times 10^{-5}$, Linsky et al.~(\citealp{linsky05})
recently proposed a \dshpe\ ratio higher than $2\times 10^{-5}$
assuming that deuterium could be significantly depleted onto dust
grains (see Jura~\citealp{jura82}; Wood et al.~\citealp{wood04};
Draine~\citealp{draine04}).

Both \dshpe\ values may challenge deuterium evolution models, the
baryonic density of the Universe inferred from primordial D/H
measurements, and our understanding of the physics of the interstellar
medium.  Regardless of the actual scenario, it is now clear that
local measurements are not enough to assess \dshpe, and that deuterium
studies in the distant interstellar medium are mandatory. Only a few
are available now.

Here we report the measurement of the deuterium abundance in the
interstellar medium toward \hd9\ (\S~\ref{hd90087}). This sight line
is studied thanks to new \fuse\ observations of this target. {\it IUE}
({\it International Ultraviolet Explorer}) archival spectra are also
used to constrain the neutral hydrogen column density from the fits of
the \lya\ line wings. Both in terms of distance and column densities,
\hd9\ provides the farthest Galactic line of sight for which deuterium
abundance has been measured from far ultraviolet absorption lines to
date.  Hence, many interstellar clouds are probed along this sight
line.  This will tend to average out the peculiarities of individual
Galactic regions, such as the Local Bubble, in order to reach
measurements that are characteristic of the interstellar medium on
large scales.

We supplement our study with a reconsideration of the \ion{O}{1}
column density measurement toward \feige\ from \fuse\ spectra,
allowing an updated interstellar D/O measurement on this sight line
(\S~\ref{feige}).

\section{The \hd9\ line of sight}
\label{hd90087}

\subsection{The target}

\hd9\ is a bright O-type star located in the fourth quadrant 
of the Galactic plane.  It was assigned different spectral types and
luminosity classes in the past, ranging from O9 to B3 and from II to
V. The most recent classification is the one given by
Mathys~(\citealp{mathys88}): O9.5III.  Savage, Meade, \&
Sembach~(\citealp{savage01}) reported the spectroscopic parallax
distance $d=2740$~pc, with an accuracy estimated to be about 30~\%
(i.e. about 800~pc here).  This value is in good agreement with the
one reported by Diplas \& Savage~(\citealp{diplas94}), $d=2716$~pc.
Table~\ref{table_star_parameters} summarizes relevant sight line and
atmospheric parameters for \hd9. This target is well beyond the Local
Bubble.  The column densities that we report below for \hd9\ represent
50-100 times the column densities usually measured within the Local
Bubble (e.g., Moos et al.~\citealp{moos02}).  The interstellar matter
located within the Local Bubble is a negligible fraction of the total
column of material probed along this sight line, which is dominated by
the distant interstellar~medium.

\subsection{Observations and data processing}

The target was first observed with \fuse\ using the low-resolution
(LWRS) aperture in April 2000 as a part of a wide study of the hot gas
in the Milky Way through the \ion{O}{6} doublet observations (Bowen et
al.~\citealp{bowen04}). Several \ion{D}{1} interstellar absorption
lines were detected against the stellar continuum on this 3.9~ksec
snapshot observation.  This indicated that the target is well-suited
for distant deuterium abundance measurement. So, an additional 14~ksec
observation was performed in June 2003 through the medium-resolution
(MDRS) aperture, allowing higher signal-to-noise ratio to be reached
and reducing interference from geocoronal emission from the
terrestrial atmosphere.  The analysis presented below uses both the
LWRS and MDRS data.  The two observations were obtained in histogram
\fuse\ mode. The observing log is summarized in Table~\ref{table_obslog}. 
Details of the \fuse\ instrument may be found in Moos et
al.~(\citealp{moos00}) and Sahnow et al.~(\citealp{sahnow00}).  Note
that the LWRS spectra of \hd9\ were first reported in the atlas of
Galactic OB \fuse\ spectra presented by Pellerin et
al.~(\citealp{pellerin02}).

The one-dimensional spectra were extracted from the two-dimensional
detector images and calibrated using version 2.4.1 of the CALFUSE
pipeline.  The data from each channel and segment (SiC1A, SiC2B,
LiF1A, LiF2B, etc.) were co-added separately for each of the two
apertures, after wavelength shift corrections of the individual
calibrated exposures.  We did not sum them because the line spread
function (LSF) and the distortions in the wavelength scale are
different for the LWRS and MDRS apertures, and even between segments
for the same aperture. We fit these datasets simultaneously in the
analysis reported below.

For the LWRS spectra, we used the 5 available exposures.  In the case
of MDRS, the individual spectra present flux variations due to the
motions of the target in the aperture, whose spread is larger that the
size of the aperture.

We did not include the exposures which have too little flux to be
useful, which led us to exclude one or two exposures from the sum of
some of the segments. Exceptions are the SiC2 segments, for which we
used only 20 of the 32 available exposures; the 12 remaining ones
present too little flux, or even no flux at all in certain cases
(because the target was out of the aperture during the entire
exposure).  Because of their good photometric quality, we use the LWRS
spectra to set the absolute flux level of the MDRS spectra.

The spectral resolution in the final spectra ranges between
$\sim13000$ and $\sim18500$, depending on detector segment and
wavelength.  The final LWRS spectra are plotted in
Fig.~\ref{fig_spectre_all}.  A sample of the SiC1B MDRS spectrum is
shown in Fig.~\ref{fig_spectre_lines} with the \ion{H}{1},
\ion{D}{1}, and \ion{O}{1} lines identified. In that figure, the flux 
has been multiplied by a factor 1.4 in order to match the LWRS one.

\  

\subsection{Data analysis}

\subsubsection{Overview}
\label{Overview}

We measured the column densities $N$ for several species on this line
of sight by fitting Voigt profiles to interstellar absorption lines.
We used the profile fitting method presented in detail by H\'ebrard et
al.~(\citealp{hebrard02}), which is based on the procedure \ow,
developed by Martin Lemoine and the French \fuse\ Team (Lemoine et
al.~\citealp{lemoine02}). We split each spectrum into a series of
small sub-spectra centered on absorption lines, and fitted them
simultaneously with Voigt profiles using \kid\ minimization (see
Fig.~\ref{fig_fit}). Each fit includes typically one hundred or more
spectral windows, and around 300 transitions of neutral atoms, ion, and
molecules. The different species are assumed to be in six different
clouds: the first one includes \ion{H}{1} only, the second one
\ion{D}{1}, \ion{O}{1}, \ion{N}{1}, \ion{Fe}{2}, and \ion{P}{2},
the third one \ion{C}{1} and its two excited levels \ion{C}{1*} and
\ion{C}{1**}, the fourth one all the rotational levels of H$_2$, the
fifth one HD, and the last one all the rotational levels of CO.

\begin{figure}
\begin{center}
\psfig{file=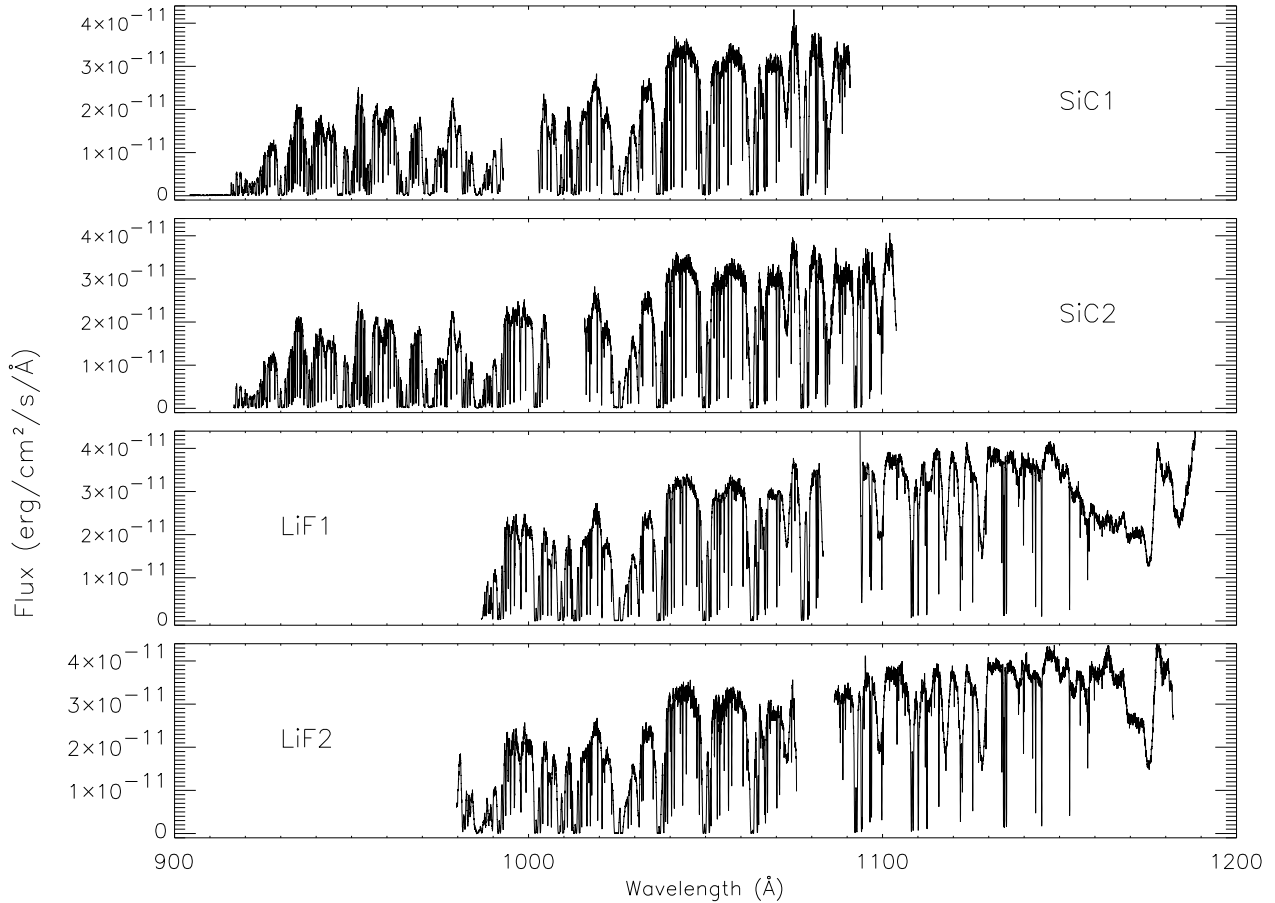,height=12.3cm}
\caption{The eight segments of the \fuse\ \hd9\ spectra obtained 
through the large aperture (LWRS).  Each channel (SiC1, SiC2, LiF1,
and LiF2) is divided in two segments (A and B), separated by a
gap. Due to the shorter exposure time, the signal-to-noise ratio is
lower for this spectrum than for the spectra obtained through the
medium aperture (MDRS), on which most of the analysis is based. The
flux calibration is however expected to be better for the LWRS
spectra.
\label{fig_spectre_all}}
\end{center}
\end{figure}

\begin{figure}
\begin{center}
\psfig{file=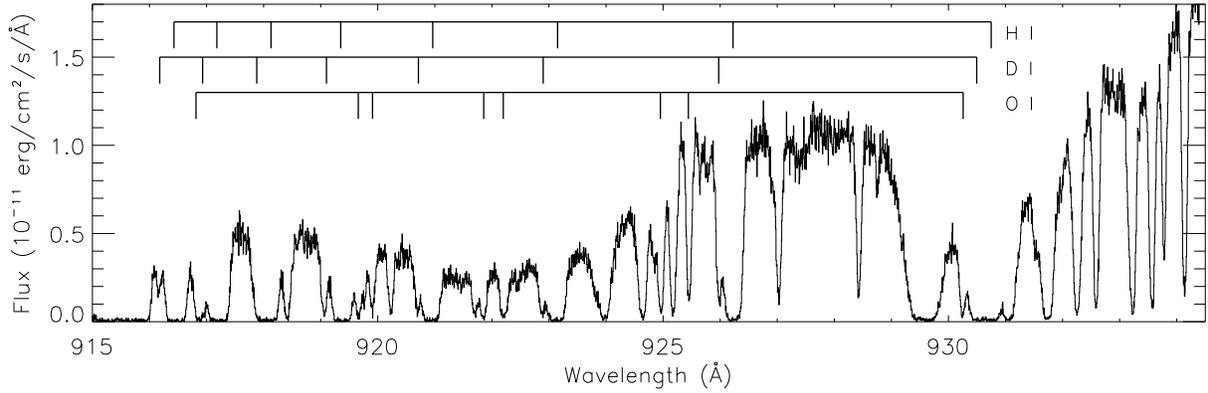,height=5.6cm}
\caption{Sample of the \fuse\ SiC1B spectrum of \hd9\ observed through
the MDRS aperture. Numerous interstellar lines are detected in 
absorption on the stellar continuum. 
Positions of \ion{H}{1}, \ion{D}{1}, and \ion{O}{1}
interstellar lines are indicated. Most of the remaining
absorption lines are H$_2$ transitions for different $J$-levels.
\label{fig_spectre_lines}}
\end{center}
\end{figure}

\begin{figure}
\begin{center}
\psfig{file=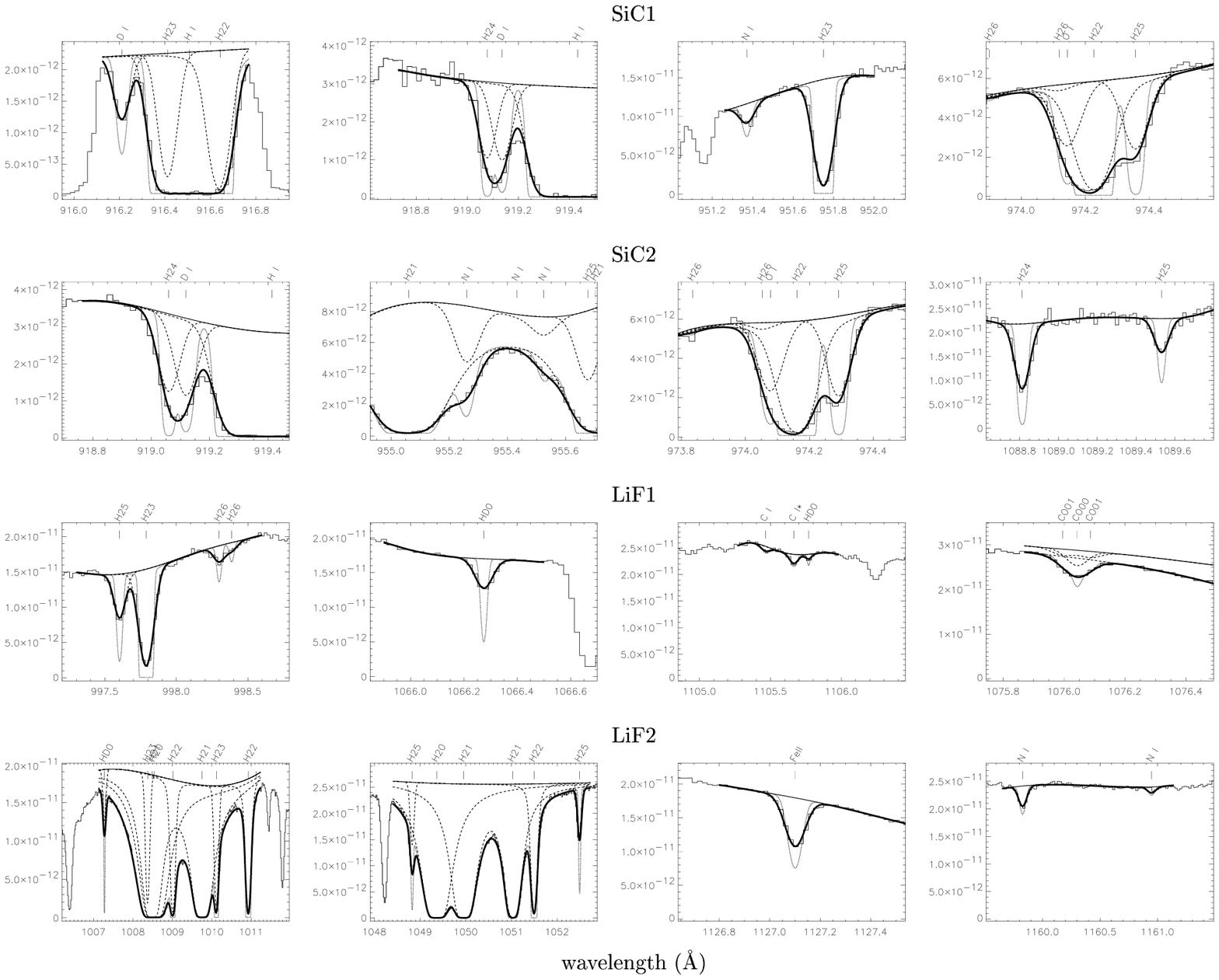,height=19cm}
\vspace{-3.5cm}
\caption{Examples of \fuse\ spectral windows fitted on the line 
of sight of \hd9.  Histogram lines are the data, the solid lines are
the fits (thick) and continua (thin). The dashed lines are the fits
for each species. The dotted lines are the model profiles prior to
convolution with the \lsf.  The Y-axis is flux in erg/cm$^2$/s/\AA.
The species are identified at the top of the plots for each line.  The
H$_2$ lines of the levels $J=0$ to $J=6$ are noted H20 to H26, HD
($J=0$) is noted HD0, and CO $J=0,1$ are noted CO00, CO01.  We show
here only 16 fitting windows for MDRS spectra, including 59
transitions. The complete fits include typically one hundred or more
spectral windows on MDRS and LWRS spectra, and around 300~transitions.
\label{fig_fit}}
\end{center}
\end{figure}

Note that thanks to the redundancy of \fuse\ spectral coverage, a
given transition might be observed in one, two, three, or four
different segments (see Fig.~\ref{fig_spectre_all}). In addition, as
LWRS and MDRS data are available for a given spectral feature, we can
have up to eight independent spectra. These different observations
allow for some instrumental artifacts to be identified and possibly
averaged out.

The laboratory wavelengths and oscillator strengths ($f$-values) used
in the analysis are from Abgrall et al.~(\citealp{abgrall93a},
\citealp{abgrall93b}) and E. Roueff (private communication) for the
molecules, and from Morton~(\citealp{morton03}) for atoms and ions.
Morton~(\citealp{morton91}) was highly used in the past in the
literature for similar studies.  Following the publication of that
atomic data compilation, some important $f$-value revisions occurred
(Beideck et al.~\citealp{beideck94}; Verner el al.~\citealp{verner94};
Tripp et al.~\citealp{tripp96}; Savage \& Sembach~\citealp{savage96};
Howk et al.~\citealp{howk00}; Sofia et al.~\citealp{sofia00}), but for
the most part, differences between Morton (1991) and the updated
atomic tables from Morton~(\citealp{morton03}) are small.  We point
out, however, the case of \ion{N}{1} transitions, for which some
oscillator strength values have been significantly updated in
Morton~(\citealp{morton03}).  Such differences (up to a few tens of
percent) would have significant effect on the \ion{N}{1} column
density measurements. Our analysis takes into account the updated
$f$-values compiled by Morton~(\citealp{morton03}), including those of
\ion{N}{1}.

Several parameters are free to vary during the fitting procedure,
including the column densities of all the species, the radial
velocities of the six interstellar clouds assumed in the fit, their
temperatures and turbulent velocities, and the shapes of the stellar
continua, which are modeled by low order polynomials. \ow\ produces
solutions that are coherent between all the fitted lines, assuming a
given cloud has a single radial velocity, temperature and turbulence.
Some instrumental parameters are also free to vary, including the flux
background, the spectral shifts between the different spectral
windows, and the widths of the Gaussian
\lsf\ used to convolve with the Voigt profiles.

\begin{figure}
\begin{center}
\psfig{file=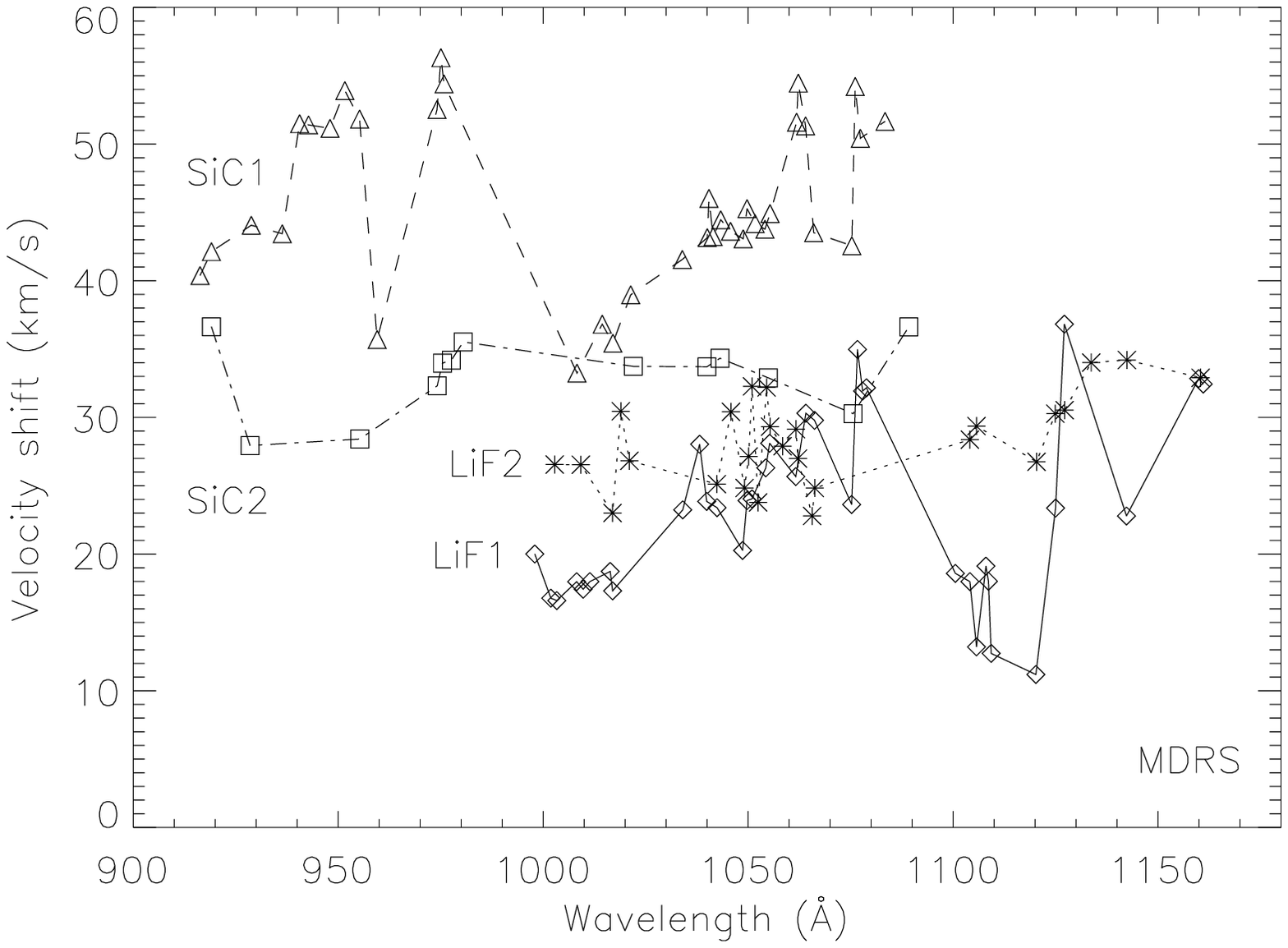,height=5.8cm}
\psfig{file=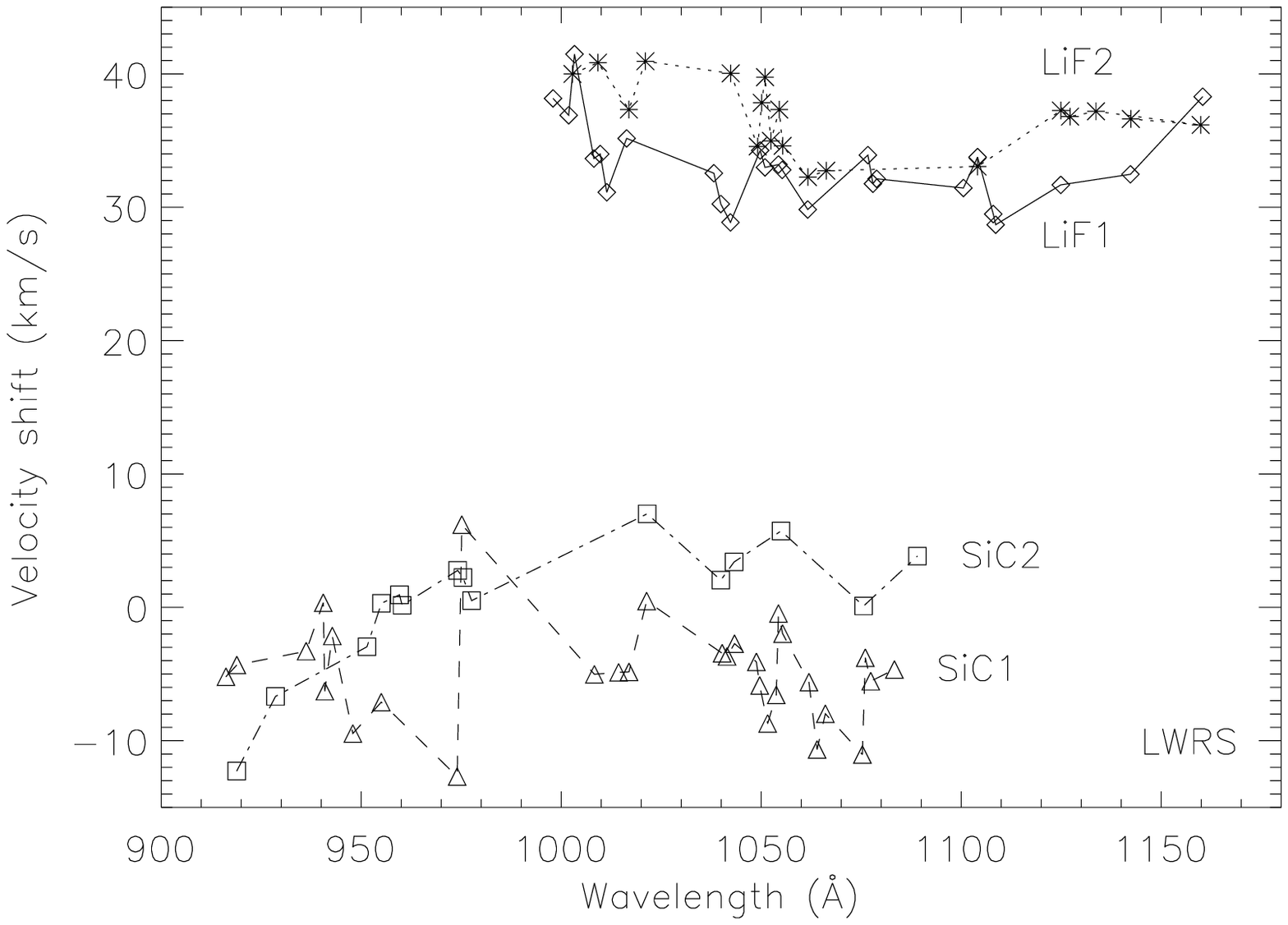,height=5.8cm}
\caption{Relative velocity shifts fitted for spectral windows on 
\hd9\ \fuse\ spectra from MDRS (left) and LWRS (right)~apertures.
\label{fig_shift}}
\end{center}
\end{figure}

\begin{figure}
\begin{center}
\psfig{file=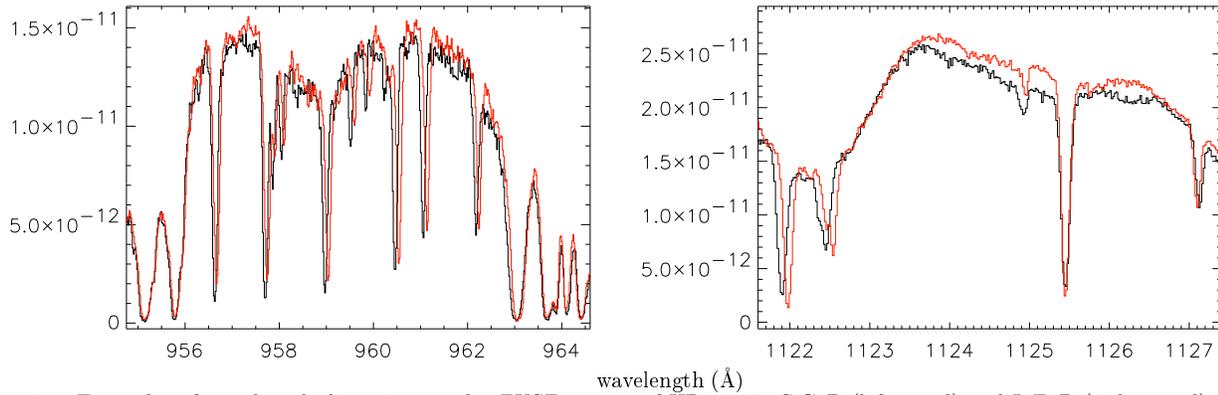,height=25cm}
\vspace{-18cm}
\caption{Examples of wavelength distortion on the \fuse\ spectra of \hd9.
SiC1B (left panel) and LiF1B (right panel) are in black, and SiC2A
(left panel) and LiF2A (right panel) are in red. The Y-axis is flux in
erg/cm$^2$/s/\AA.
\label{fig_shift_ex}}
\end{center}
\end{figure}

The velocity shifts fitted for the spectral windows are plotted in
Fig.~\ref{fig_shift}. They allow the inaccuracies of the \fuse\
wavelength calibration to be corrected.  Typically, the velocity
corrections are consistent within a few \kms\ within a given segment.
They are not a smooth function of wavelength, and shifts as high as 10
to 20~\kms\ might occur over only a few \AA ngstr\"oms in extreme
cases (Fig.~\ref{fig_shift_ex}). Larger velocity corrections might
occur from one segment to another.  The velocity shifts are relative
and we did not attempt to measure absolute (e.g. heliocentric or LSR)
velocities.  We note that the wavelength distortion problem is
generally more severe in \fuse\ data obtained in histogram mode.
Since most \fuse\ observations employ the time-tag mode, distortion of
the wavelength scale is usually not as bad as the examples shown in
Fig.~\ref{fig_shift_ex} (which were recorded in histogram~mode).

\begin{figure}
\begin{center}
\psfig{file=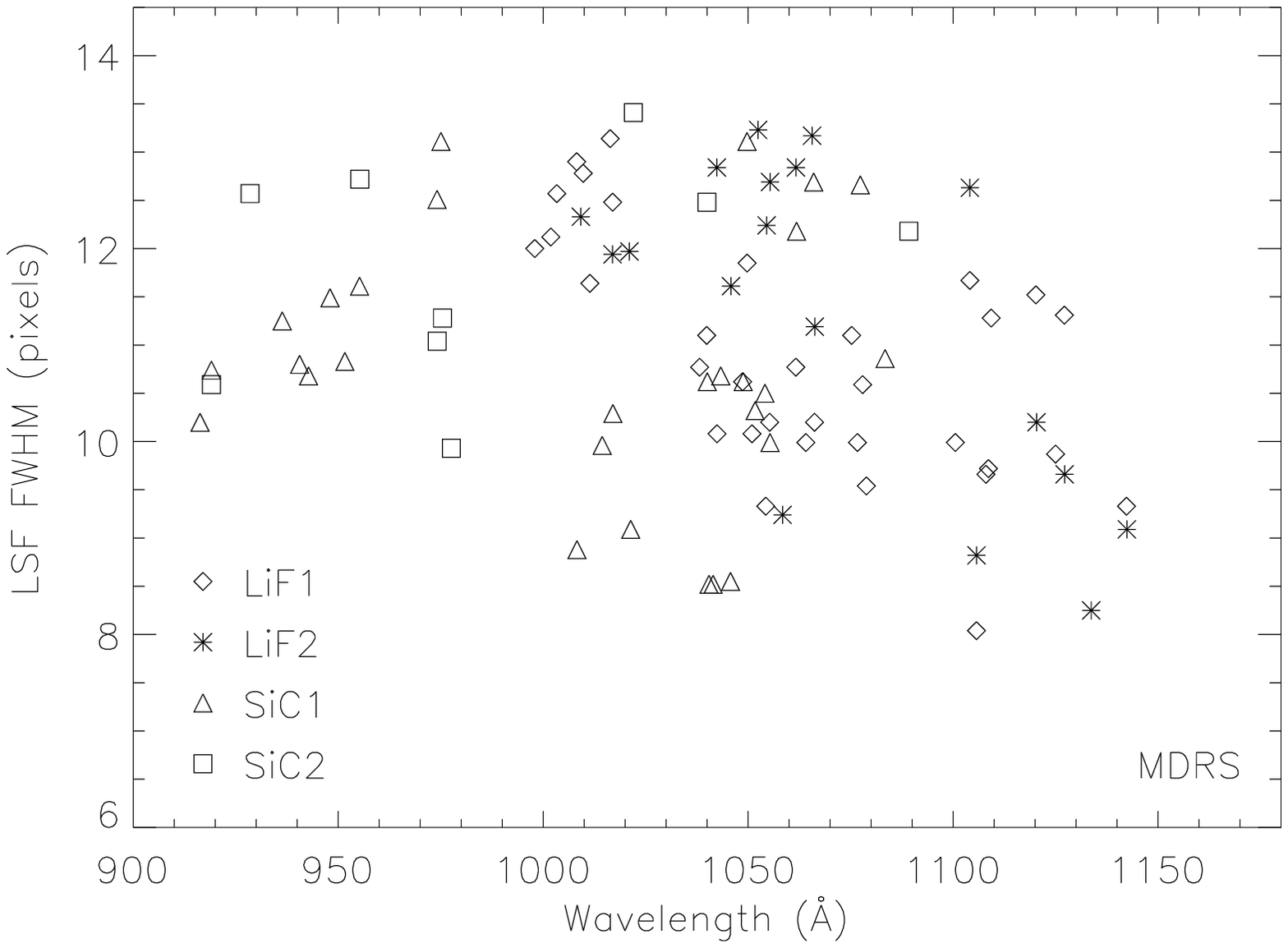,height=5.8cm}
\psfig{file=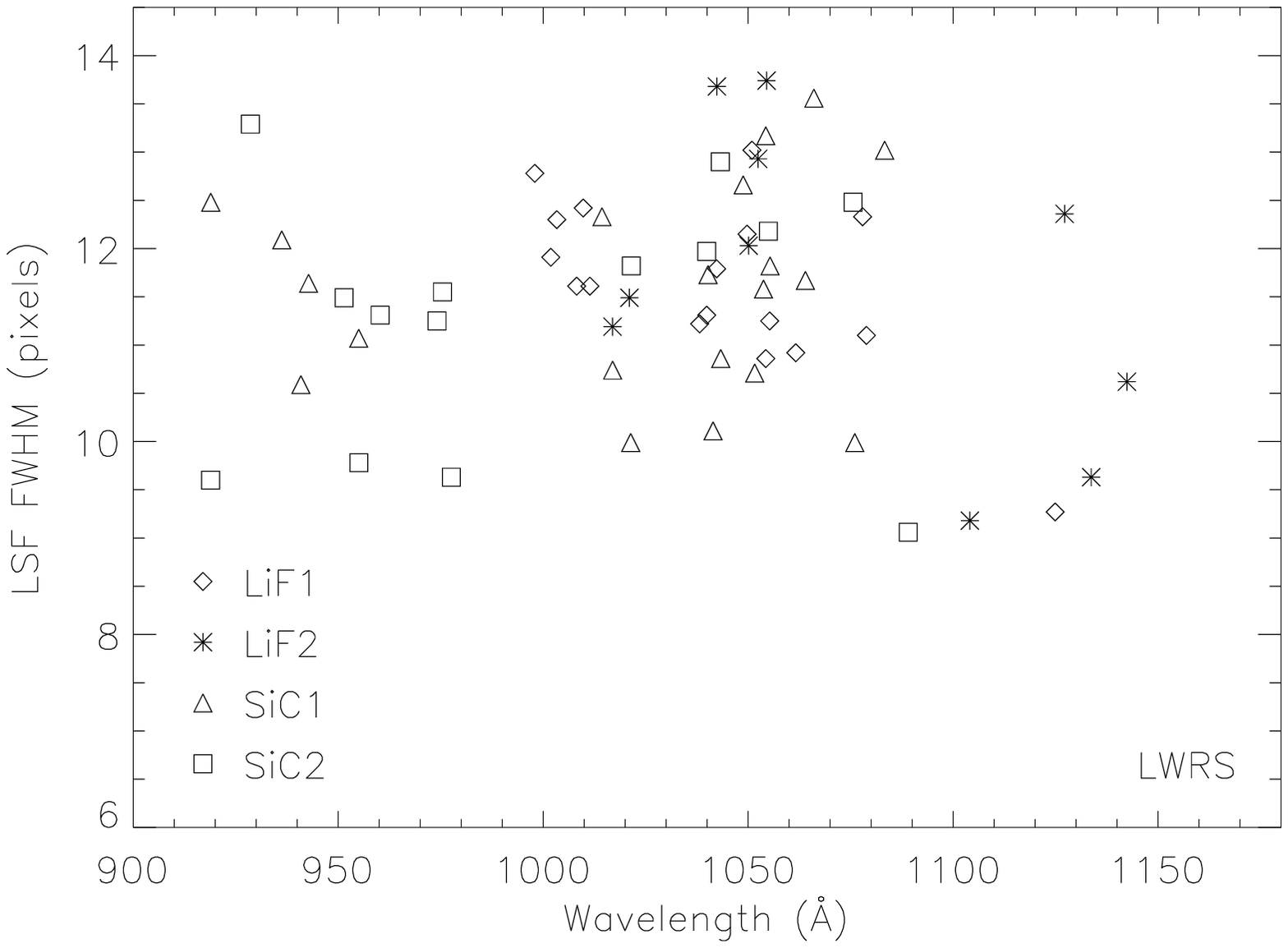,height=5.8cm}
\caption{Full widths at half maximum (FWHM) of the Gaussian \lsf\ 
(LSF) fitted for spectral windows on \hd9\ \fuse\ spectra from MDRS
(left) and LWRS (right)~apertures.
\label{fig_fwhm}}
\end{center}
\end{figure}

The widths of the Gaussian \lsf\ fitted for the spectral windows are
shown in Fig.~\ref{fig_fwhm}. The averages for the apertures LWRS and
MDRS are respectively $11.6\pm1.2$ and $11.0\pm1.4$ \fuse\ pixels
(full widths at half maximum). We also attempted to fit all spectral
windows with a unique LSF width. The best \kid\ was obtained for a
11.4-pixel width.  However, significant variations occur from one
segment to the other, and as a function of the wavelength for a given
segment. These averaged widths are in agreement with those reported in
H\'ebrard et al.~(\citealp{hebrard02}). Note that during the fits, all
the spectra were binned to three \fuse\ pixel samples in order to
reduce computing time. Considering the width of the LSF, this binning
does not degrade the spectral resolution.

When possible, the saturated lines located on the flat part of the
curve of growth were excluded from the fit. Indeed, such saturated
lines can introduce systematic effects on column density measurement
(H\'ebrard et al.~\citealp{hebrard02}; H\'ebrard \&
Moos~\citealp{hebrard03}; Friedman et al.~\citealp{friedman05}),
mainly because the actual widths of the lines, the velocity structure
of the sight line, or the \lsf\ of the instrument are unknown but
expected to be complex.  A particular difficulty with distant lines of
sight such as \hd9\ is the rarity of unsaturated lines for some
species, especially \ion{D}{1} and \ion{O}{1}, which is caused by high
column densities.  This is significantly different from local sight
lines, for which numerous unsaturated lines are~available.

Another relevant difference between the long line of sight toward
\hd9\ and those of targets located in the Local Bubble is the 
number of detected absorption lines. The overview of the \fuse\ \hd9\
spectra plotted in Fig.~\ref{fig_spectre_all} can be compared with the
Fig.~1 in H\'ebrard et al.~(\citealp{hebrard02}), which shows the
corresponding \fuse\ spectra for a target located 53~pc away. The
\hd9\ spectra are particularly crowded. This is due to stellar lines,
which are more numerous here than in the case of the targets used to
probe the more local interstellar medium (mainly white dwarfs). In the
case of \hd9, however, the high rotational velocity of this star
smooths the stellar continuum. An even bigger effect on the absorption
line number arises from interstellar lines, as the column densities
here are typically tens or hundreds of times larger than those
measured in the Local Bubble.  In particular, many H$_2$ lines are
detected in the \hd9\ spectra. This makes line blending a critical
problem, which can be resolved with simultaneous fits of all the lines
together.

Only one interstellar component, at a given radial velocity, was
assumed for the sight line for a given species.  This assumption is
unlikely to be true but it will have no effect on the column densities
measured from unsaturated lines (see tests of this assumption in
H\'ebrard et al.~\citealp{hebrard02}).  Thus, we report total column
densities, integrated along each line of~sight. 

The simultaneous fit of numerous lines allows the reduction of
statistical and systematic errors, especially those errors due to
continua placement, LSF uncertainties, line blending, flux and
wavelength calibrations, and atomic data uncertainties. The error bars
were obtained using the \deltakid\ method presented by H\'ebrard et
al.~(\citealp{hebrard02}).  We also estimated the error bars by
varying the assumptions for the fits.  This included the LSF (normally
free to vary, but tests were made with fixed LSFs), the velocity
structure (several species are assumed to be in the same component,
but fits with one component per species were performed) and also the
aperture (comparing (a) MDRS and LWRS together, (b) only LWRS, and (c)
only~MDRS).

\begin{figure}
\begin{center}
\psfig{file=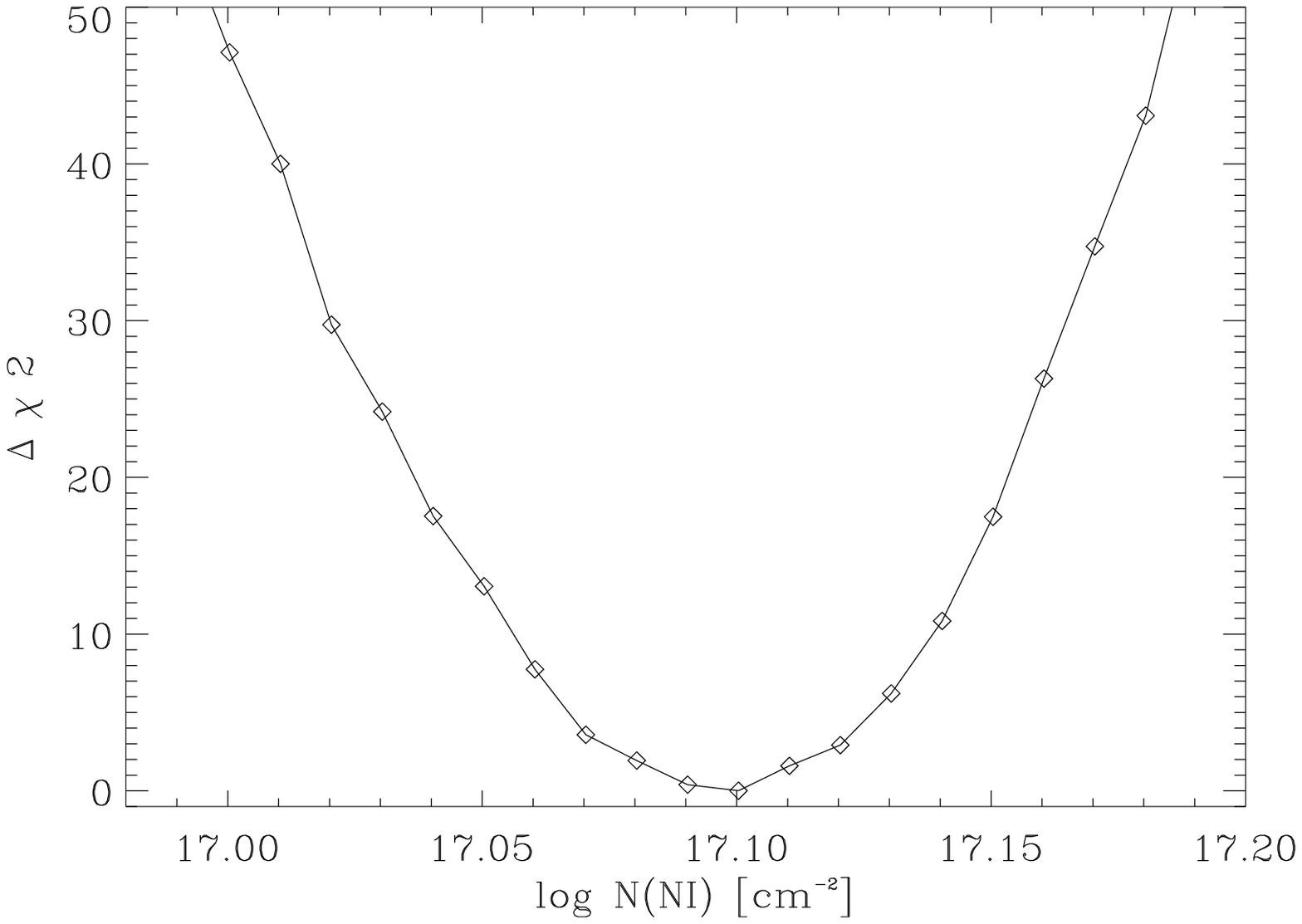,height=4.35cm}
\psfig{file=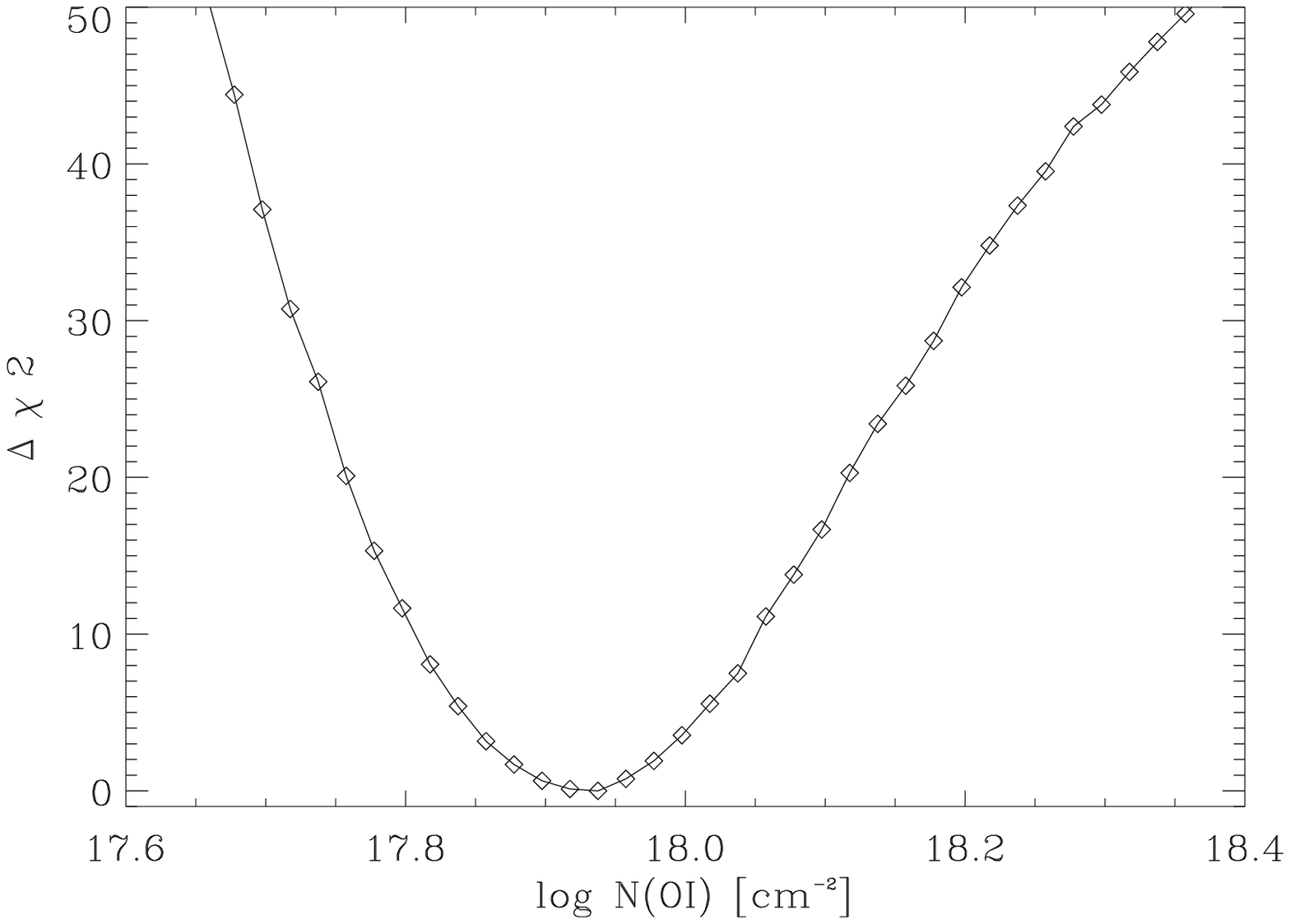,height=4.35cm}
\psfig{file=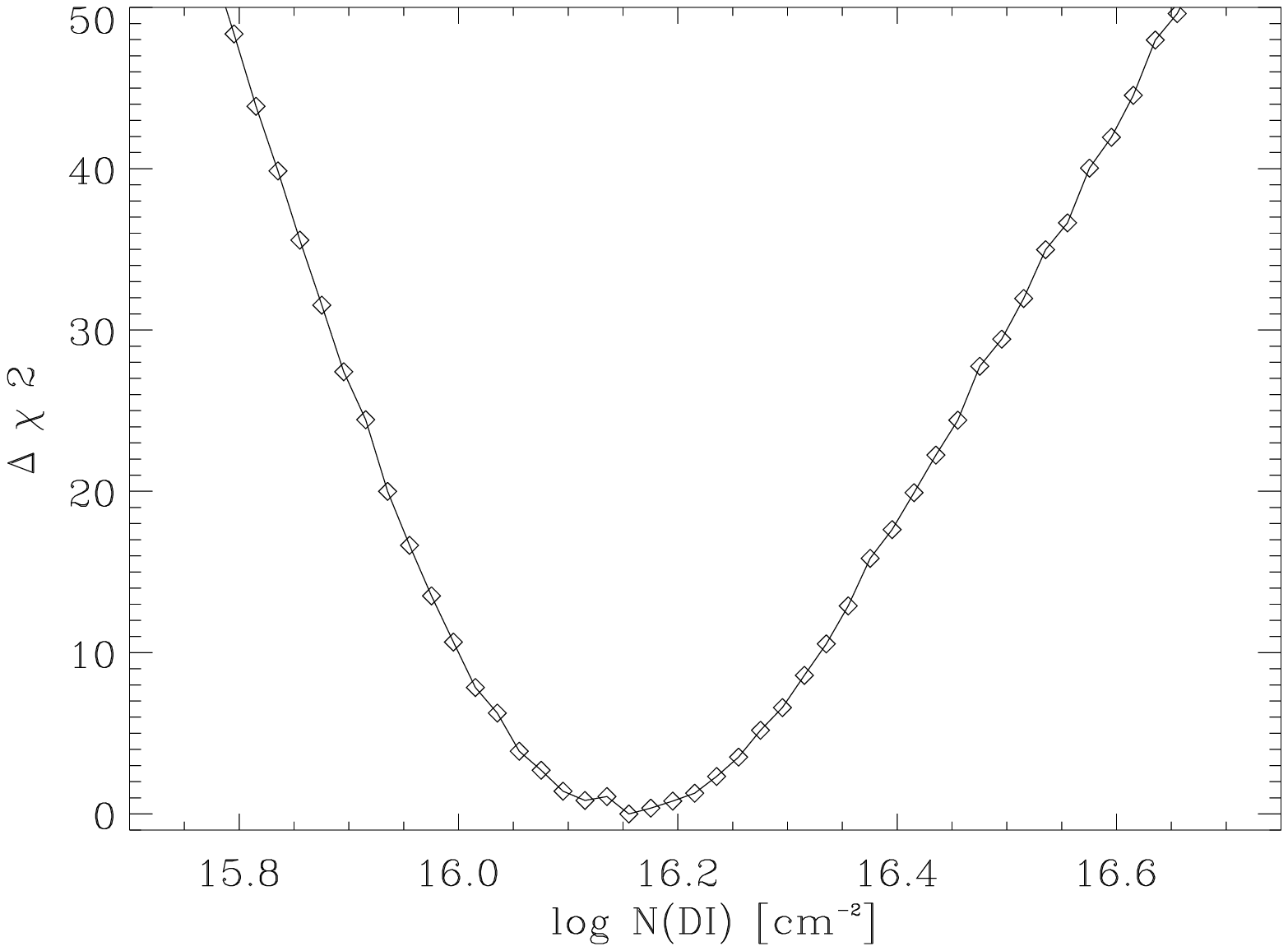,height=4.35cm}
\caption{\deltakid\ curves for \ion{N}{1}, \ion{O}{1}, and
\ion{D}{1} toward \hd9. 
Each point on these curves corresponds to an individual fit, made 
with all the parameters free, but the \ion{N}{1}, \ion{O}{1}, or 
\ion{D}{1} column density fixed. 
\kid\ values are rescaled on these plots (see H\'ebrard 
et al.~\citealp{hebrard02}).
\label{fig_chi2_DON}
}
\end{center}
\end{figure}

We discuss below the different species. The measured column densities
and error bars are reported in Table~\ref{table_columns} and examples
of fits are given in Fig.~\ref{fig_fit}. The \deltakid\ curves for
\ion{N}{1}, \ion{O}{1}, and \ion{D}{1} are plotted in 
Fig.~\ref{fig_chi2_DON}.

\subsubsection{Molecular hydrogen (H$_2$)}
\label{H2}

All the rotational levels of H$_2$ were included in a single
interstellar component.  The fits used a total of about $120$ H$_2$
absorption lines from rotational levels $J=0$ to $J=6$ available in
one to eight independent segment/aperture configurations. The $J=0,1$
H$_2$ lines present damping wings and the $J=4,5,6$ that we used are
unsaturated. This allows accurate column density measurements to be
performed for these five levels. Only saturated lines located on the
flat part of the curve of growth are available for the $J=2,3$ H$_2$
transitions; this precludes accurate measurements of their column
densities.

\begin{figure}
\begin{center}
\psfig{file=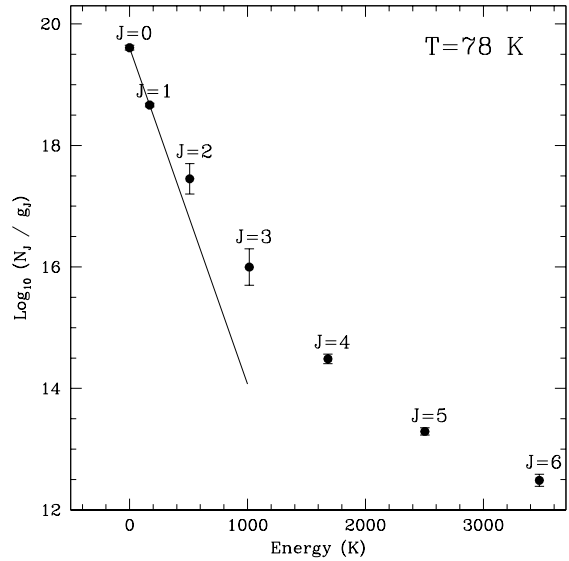,height=8cm}
\caption{Excitation diagram of the H$_2$ lines detected toward \hd9. 
The column densities of the rotational levels $J=0$ and $J=1$ are
consistent with a temperature $T = 78$~K.
\label{fig_excitation}}
\end{center}
\end{figure}

The excitation diagram corresponding to the derived H$_2$ column
densities is shown in Fig.~\ref{fig_excitation}. H$_2$ is not
thermalized for levels $J\ge2$. The column densities of the levels
$J=0$ and $J=1$ give the temperature $T=78$~K, which can be considered
as the kinetic temperature of the H$_2$ molecular gas.  A temperature
of 78~K would correspond to $b=0.8$~\kms.  Thus, the single-component
Doppler parameter $b\simeq7$~\kms\ that we found for this molecular
component is not purely thermal but contains additional contributions
from the gas turbulent motions and/or additional velocity
components. Note that since the $J=0,1$ levels are damped and the
$J=4$ to $6$ levels are optically thin, the Doppler parameter we
derived is mainly characteristic of the $J=2,3$ levels. On another
hand, since we did not detect any significant velocity shift between
the different rotational levels (all are $< 1$~\kms), this suggests
that these levels arise from the same dominant cloud component(s) on
average.

\subsubsection{Deuterated molecular hydrogen (HD)}
\label{HD}

The only HD lines detected in the spectra are those corresponding to
the level $J=0$; no HD lines are detected for $J\ge1$.  Our global
fits include the seven HD lines used by Lacour et
al.~(\citealp{lacour05}), and also four extra ones at 1007.28~\AA\
(see Fig.~\ref{fig_fit}, left-bottom panel), 925.78~\AA, 967.53~\AA,
and 1042.84~\AA.  The other HD ($J=0$) lines are blended with strong
lines (as $\lambda$\,1001.89\,\AA\ and $\lambda$\,1078.81\,\AA, which
are lost in saturated H$_2$ lines) and are not detected.  Among those
eleven HD lines, the five strongest appear to be saturated or nearly
so.  We then performed extra fits including only the six weakest HD
lines, namely $\lambda$\,1054.29\,\AA, $\lambda$\,959.82\,\AA,
$\lambda$\,967.53\,\AA, $\lambda$\,925.78\,\AA,
$\lambda$\,1066.27\,\AA, and $\lambda$\,1105.83\,\AA\ (in decreasing
order of $f$-values). Excluding those five saturated lines increased
the HD column density estimate by 0.1~dex, from 14.42 to 14.52.  We
adopt the latter value.

Interestingly, the radial velocity of the HD lines was found to be
systematically redshifted by $4\pm0.5$~\kms\ with respect to the
velocity of the H$_2$ lines. According to the error bars, the velocity
shift is significant; it is mainly constrained by the 13 spectral
windows that include close HD and H$_2$ lines.  By comparison, no
velocity shifts were detected between the different rotation levels of
H$_2$ (\S~\ref{H2}), between the different atomic or ionic species
(\S~\ref{NI}), nor between \ion{H}{1} and \ion{D}{1}
(\S~\ref{HI}). Thus, the shift between H$_2$ and HD appears to be an
actual effect, and not a systematic effect due to an instrumental
artifact.  In addition, we found a Doppler parameter for the HD lines
of $b\simeq3.5$~\kms, about half the Doppler parameter estimated for
H$_2$ (mainly constrained by $J=2,3$).  These two results therefore
suggest that HD and H$_2$ rotational lines do not trace exactly the
same gas conditions.  Consequently, the HD and H$_2$ lines were not
assumed to be located in the same component during the fits.  This
might jeopardize the interpretation of the HD/2H$_2$ ratio computed
from the measured column densities of the two species. Note that a
similar radial velocity shift (4.6~\kms) was also detected between
H$_2$ and HD on the sight line to PG\,0038+199 (Williger et
al.~\citealp{williger05}). As numerous different H$_2$ and HD
transitions are used in these two analyses, this suggests that HD
tabulated wavelengths are unlikely to be the cause of the velocity
shifts. A broad study of H$_2$ and HD radial velocities toward
numerous targets is however mandatory to progress on this~issue.

\subsubsection{Nitrogen (\ion{N}{1})}
\label{NI}

Numerous unsaturated \ion{N}{1} lines are detected in the \fuse\
spectra of \hd9.  The fits include eight of them, namely
$\lambda$\,955.26\,\AA, $\lambda$\,959.49\,\AA,
$\lambda$\,951.29\,\AA, $\lambda$\,955.53\,\AA, 
$\lambda$\,960.20\,\AA, $\lambda$\,1159.82\,\AA,
$\lambda$\,1160.94\,\AA, and $\lambda$\,955.44\,\AA\ (in decreasing
order of $f$-values). Some of these lines are detected in several
segments and/or through the two available apertures. Thus, the final
fits include 18 independent unsaturated \ion{N}{1} lines. These lines
were fitted simultaneously with all the other ones, assuming that they
are in the same cloud as \ion{D}{1}, \ion{O}{1}, \ion{Fe}{2}, and
\ion{P}{2}.  Indeed, these species are supposed to co-exist in neutral
regions (\ion{Fe}{1} and \ion{P}{1} ionization potentials are
respectively 7.9~eV and 10.5~eV, i.e. below the hydrogen ionization
potential). We checked this hypothesis by performing fits with these
five species in five different components, and we did not detect any
significant radial velocity shifts between them.  The final error bar
reported in Table~\ref{table_columns} for $N$(\ion{N}{1}) is small due
to the numerous unsaturated lines available.

\subsubsection{Oxygen (\ion{O}{1})}
\label{hd900987_OI}

Many \ion{O}{1} lines are available in the \fuse\ bandpass. However,
in the range $\log N($\ion{O}{1}$)\simeq17-18$, most of those lines
are saturated, which precludes accurate column density measurement (see
\S~\ref{feige} and Friedman et al.~\citealp{friedman05}).  In this
column density range, only one unsaturated \ion{O}{1} line is
available in the \fuse\ bandpass, namely $\lambda\,974.07\,$\AA. This
weak line is blended with three H$_2$ lines, including two lines
(levels $J=2$ and $J=5$) with equivalent widths larger than the
equivalent width of $\lambda\,974\,$\AA. The third one (level $J=6$)
is negligible.  The H$_2$ contributions to this blend can be
calculated thanks to the $\sim120$ H$_2$ lines included in the fits,
many of which are unblended.  Thus, the shapes of the three
$\lambda\,974\,$\AA\ H$_2$ lines are~well-constrained.

$\lambda\,974\,$\AA\ is detected in the segments SiC1B and SiC2A,
through the two available apertures. Our fits include these four
different observations of this line, which reduces potential
instrumental effects.  The fits of $\lambda\,974\,$\AA\ observed
through MDRS on SiC1B and SiC2A are plotted in Fig.~\ref{fig_fit}. The
H$_2$ ($J=5$) line seems slightly under-fitted, which might be due to
a weak unknown feature.  However, this extra absorption appears to be
weak enough and far enough from $\lambda\,974\,$\AA\ to have no
significant effect on~$N($\ion{O}{1}$)$.

\begin{figure}
\begin{center}
\psfig{file=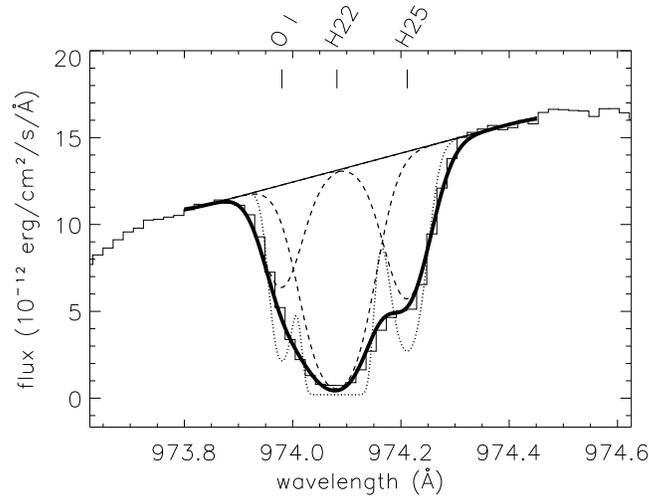,height=8cm,angle=90}
\vspace{1cm}
\caption{Fit of the $\lambda\,974.07\,$\AA\ \ion{O}{1} line as observed 
for \fuse\ spectra of HD\,195965.  Same conventions as in
Fig.~\ref{fig_fit}.
The value $N($\ion{O}{1}$)=17.77\pm0.06$ obtained from this fit is in
good agreement with those obtained by Hoopes et al.~(2003) using the
\ion{O}{1} line at $1355.60\,$\AA, namely $17.76\pm0.06$. Apparently,
$\lambda\,974\,$\AA\ presents no strong oscillator strength
inconsistencies nor significant unresolved blends.
\label{fig_hd195}
}
\end{center}
\end{figure}

Measuring a column density from only one transition makes the result
highly sensitive to possible errors caused by an erroneous $f$-value
or uncontrolled blends from unknown lines. However, up to now, no
clues suggest such problems. No strong inconsistencies were found
between $\lambda\,974\,$\AA\ and the stronger \ion{O}{1} lines
available in the \fuse\ bandpass in the case of
BD$\,$+28$^{\circ}$4211 (H\'ebrard \&
Moos~\citealp{hebrard03}). Slight disagreements might be difficult to
see, however, as $\lambda\,974\,$\AA\ is weak on this sight line. We
also measured \ion{O}{1} in the sight line of HD\,195965 from \fuse\
data, using $\lambda\,974\,$\AA\ with the same method as those we use
for \hd9\ (see Fig.~\ref{fig_hd195}).  We obtained $\log
N($\ion{O}{1}$)=17.77\pm0.06$, in agreement with the value obtained by
Hoopes et al.~(\citealp{hoopes03}) using $\lambda\,1356\,$\AA, $\log
N($\ion{O}{1}$)=17.76\pm0.06$.  Thus, there are apparently no
inconsistencies between $\lambda\,974\,$\AA\ and $\lambda\,1356\,$\AA\
(the studies of oxygen abundance by Meyer et al.~(\citealp{meyer98}),
Andr\'e et al.~(\citealp{andre03}), or Cartledge et
al.~(\citealp{cartledge04}) are based on $\lambda\,1356\,$\AA).
The two comparisons we present above suggest that $\lambda\,974\,$\AA\
presents no strong oscillator strength inconsistencies nor significant
uncontrolled~blends (see also~\S~\ref{hd900987_others}).

\subsubsection{Deuterium (\ion{D}{1})}

As in the case of \ion{O}{1} discussed in the previous subsection,
only a few unsaturated \ion{D}{1} lines are available in this high
column density regime.  Strong blending (especially for H$_2$ lines)
and saturation prohibit the use of most of the \ion{D}{1} lines.  The
\ion{D}{1} lines of $\lambda\,920.7\,$\AA, $\lambda\,922.9\,$\AA,
$\lambda\,926.0\,$\AA, and at larger wavelengths in the Lyman series
are all saturated, so we did not include them in the
fits. $\lambda\,917.9\,$\AA\ is not detected because it is located at
$\sim-10$~\kms\ from a strong H$_2$ ($J=1$) line with damping wings,
so it cannot be used to constrain $N($\ion{D}{1}$)$.  We could not
obtain a good fit to $\lambda\,916.9\,$\AA. This \ion{D}{1} line is
blended with two strong lines (\ion{O}{1} and H$_2$ ($J=3$)) but
apparently also with some unknown feature(s).

Our fit therefore includes only two \ion{D}{1} transitions, namely
$\lambda\,916.2\,$\AA\ and $\lambda\,919.1\,$\AA.  The first one is
available only on segment SiC1B, while the second one is available on
both SiC1B and SiC2A. As we use simultaneously the data from the MDRS
and LWRS apertures, our fits include a total of six \ion{D}{1} lines.

A weak H$_2$ ($J=5$) line is located $\sim-30$~\kms\ from
$\lambda\,916.2\,$\AA. It is weak and far enough to have no
significant effects. $\lambda\,916.2\,$\AA\ is probably the highest
detectable \ion{D}{1} in the Lyman series.  Indeed, in the high column
density regime necessary to produce significant absorption for Lyman
lines this high, interstellar absorption lines are too numerous and
crowded below 916~\AA, and therefore almost no flux is detected there
(see Figs.~\ref{fig_spectre_all} and \ref{fig_spectre_lines}).

The $\lambda\,919.1\,$\AA\ \ion{D}{1} line is blended with H$_2$
($J=4$), located $\sim20$~\kms\ blueward and of similar equivalent
width. The fits include nine other H$_2$ ($J=4$) lines, observed on
different segments and through the two apertures, allowing the blend 
to be deconvolved.  $\lambda\,919.1\,$\AA\ is slightly saturated.  We 
tested the results by including only $\lambda\,916.2\,$\AA, and the 
ensuing $N($\ion{D}{1}$)$ was in agreement with the fit to $\lambda
919.1$~\AA.

The choice of the continuum used for fitting the absorption lines has
an effect on the measured column density. Here the continua are fitted
by polynomials for each spectral window. As the parameters of the
polynomials are free to vary, the uncertainty due to the continuum
should be included in the final error bar. We checked that by varying
the continua, i.e. by using different polynomial degrees or changing
the wavelength intervals on which they are fitted.

Our result is $\log N($\ion{D}{1}$)=16.16\pm0.12$ (\2s). To date, this
is the largest published \ion{D}{1} column density ever measured.  As
$\lambda\,916.2\,$\AA\ would saturate for $N($\ion{D}{1}$)$ values 2
or 3 times larger, it would be difficult to accurately measure larger
\ion{D}{1} column densities for other sight lines. Note, that this limit 
also provides a selection bias against sight lines with higher D/H or
D/O ratios in this high column density regime. Interesting lower
limits on $N($\ion{D}{1}$)$ might however be estimated from saturated
\ion{D}{1} lines.

The column density of HD represents only $2.3\pm1.0$~\%\ (\2s) of
$N($\ion{D}{1}$)$. Most of the deuterium detected on the line of sight
of \hd9\ is in atomic form.  Therefore, given the size of the errors,
deuterium in molecular form can be neglected for the computation of
the total deuterium column density.

\subsubsection{Other species \\(\ion{Fe}{2}, \ion{P}{2}, \ion{Ar}{1}, 
\ion{C}{1}, \ion{C}{1*}, \ion{C}{1**}, CO)}
\label{hd900987_others}

Interstellar absorption lines of CO, \ion{Ar}{1}, \ion{Fe}{2},
\ion{P}{2}, \ion{C}{1} and its two excited levels \ion{C}{1*} and 
\ion{C}{1**} are also detected in the spectra of \hd9. We report 
accurate column density measurements for CO, \ion{Fe}{2}, and
\ion{P}{2}. 
The measurements for the remaining species are more uncertain.  We
discuss here briefly all these species.

Numerous unsaturated \ion{Fe}{2} lines are available in the \fuse\
spectra of \hd9. We used the following ones: $\lambda\,1142.4\,$\AA,
$\lambda\,1064.0\,$\AA, $\lambda\,1062.2\,$\AA,
$\lambda\,1127.1\,$\AA, and $\lambda\,1083.4\,$\AA\ (in decreasing
order of $f$-values). We also made fits including these lines and
$\lambda\,1055.3\,$\AA\ and $\lambda\,1133.7\,$\AA, whose $f$-values
are larger and which might be slightly saturated. No significant
difference was found.

For \ion{P}{2}, we used only $\lambda\,1124.9\,$\AA, since
$\lambda\,1152.8\,$\AA\ is saturated.  As \ion{Fe}{2}, the \ion{P}{2}
lines are at the same radial velocity as \ion{D}{1}, \ion{O}{1}, and
\ion{N}{1}. Lebouteiller et al.~(\citealp{lebouteiller05}) show that 
\ion{P}{2} might be considered as a good proxy for \ion{O}{1}. The 
\ion{P}{2}/\ion{O}{1} ratio we measure toward \hd9\ is in good agreement 
with the average ratio they report. This is another argument to
support the reliability of our measured $N($\ion{O}{1}$)$, based on
$\lambda\,974\,$\AA\ (see \S~\ref{hd900987_OI}).

The CO A-X bands at $1076\,$\AA\ and $1088\,$\AA\ were also detected
toward \hd9. Since the band structure cannot be resolved with the
\fuse\ data, we included $J$ levels up to $J=7$ in our initial fits.
Our best fit indicated that the column densities of the $J=0$ and $1$
rotational levels are similar while the contribution from the $J\ge2$
levels is negligible (column densities ten times lower or less). The
CO lines are detected at the same radial velocity as H$_2$.  In both
regions, there is one $J=0$ CO transition, and two $J=1$ CO
transitions on each side of the first one.  The lines are broad enough
to allow the two $J$-levels to be measured, despite their not being
resolved. The excitation diagram leads to a temperature $T_{\rm ex}$
of about 4~K, indicating that the rotational levels are sub-thermally
populated. The derived CO/H$_2$ ratio is $(6.5\pm1.4)\times10^{-7}$
(see Table~\ref{table_ratios}), a value about 3 orders of magnitude
smaller than that typically found in purely molecular clouds. The
latter results are consistent with rotational population distributions
and CO abundances typically found in diffuse molecular clouds where
photo-processes still largely dominate the gas chemical pathways
(Sonnentrucker et~al.~\citealp{sonnentrucker03}).

For carbon, the strongest \ion{C}{1} line, at 945~\AA, is near to
saturation. Numerous weaker \ion{C}{1} lines are detected in the range
$1103-1111$~\AA, together with \ion{C}{1*} and \ion{C}{1**}
transitions. However, it is not possible to obtain a fit in agreement
for the different transitions. This is probably due to inaccurate
$f$-values. Using a large sample of interstellar \ion{C}{1} lines
recorded at high resolution, Jenkins \& Tripp~(\citealp{jenkins01})
have shown that many \ion{C}{1} $f$-values require significant
revisions, especially $f$-values of weak transitions.  Unfortunately,
they did not analyze \ion{C}{1} transitions in the \fuse\ wavelength
range.  Our reported carbon column density measurements therefore
remain inaccurate.

Finally, $N$(\ion{Ar}{1}) is obtained only from saturated lines
($\lambda\,1066.7\,$\AA\ and $\lambda\,1048.2\,$\AA) on the flat part
of the curve of growth; it is thus uncertain.

\subsubsection{Hydrogen (\ion{H}{1})}
\label{HI}

\begin{figure}
\begin{center}
\psfig{file=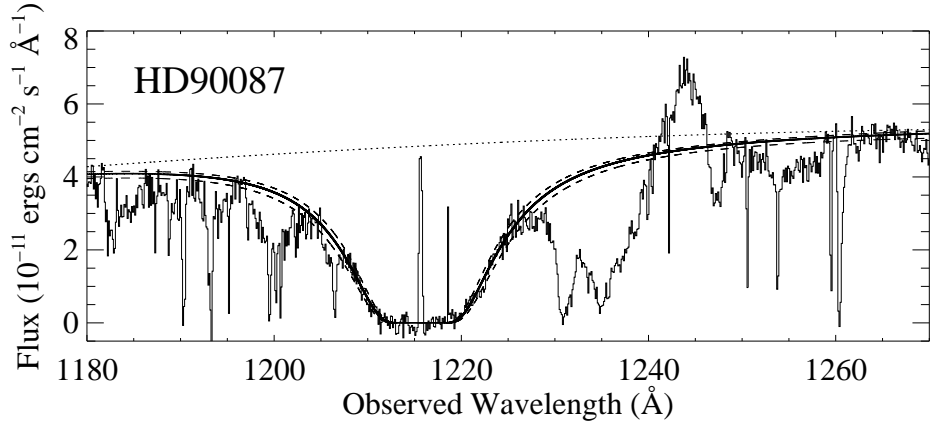,height=14cm,angle=90}
\caption{Fit of the \lya\ \ion{H}{1} line toward \hd9.
The black line is the high resolution {\it IUE} spectrum.  The strongly
damped \lya\ profile is readily apparent along with the nearby
\ion{N}{5} P\,Cyg profile ($\lambda\,1240\,$\AA) and a variety of
stellar and interstellar absorption lines.  The emission line in the
center of the \lya\ profile is the geocoronal \ion{H}{1} emission due
to the terrestrial atmosphere (the narrow spike at the right edge of
the saturated core is a spurious pixel). The solid line shows the best
fit, and the two dashed lines show the $\pm$\1s\ fits. The dotted line
in the plot shows the adopted continuum for the best fit.
\label{fig_HI_iue}
}
\end{center}
\end{figure}

To determine the total \ion{H}{1} column density in the interstellar
medium toward \hd9, we fitted the damped \lya\ profile recorded in an
{\it IUE} spectrum of the star. \hd9\ was observed with {\it IUE} once
in 1982 (Table~\ref{table_obslog}).  We retrieved the IUESIPS ({\it
IUE Spectral Image Processing System}) version of the data and
extracted the spectrum using standard
procedures. Fig.~\ref{fig_HI_iue} shows the portion of the spectrum in
the vicinity of the \lya\ line.  The strongly damped \lya\ profile is
readily apparent along with the nearby \ion{N}{5} P\,Cygni profile
(which causes the emission feature at $\sim1241$\,\AA\ and the
adjacent trough at $\sim1239$\,\AA) and a variety of stellar and
interstellar absorption lines.

We measured $N($\ion{H}{1}$)$ using the procedure described in Jenkins
et al.~(\citealp{jenkins99}) and Sonneborn et
al.~(\citealp{sonneborn00}).  Briefly, we used minimized \kid\ in
order to determine the \ion{H}{1} column that provided the best fit to
the Lorentzian wings of the \lya\ profile.  The following five
parameters were freely varied using Powell's method (Press et
al.~(\citealp{press95}): (1) $N($\ion{H}{1}$)$, (2)-(4) three
coefficients of a low-order polynomial fitted to the continuum 
over a broad range on each side of \lya, and (5) a correction for the
flux zero level.  \hd9\ is an early-type star, and any stellar \lya\ 
line is unlikely to significantly affect the interstellar \lya\ 
profile, especially since the interstellar \ion{H}{1} column toward
this star turns out to be large and the \lya\ profile is accordingly
quite broad. To evaluate upper and lower confidence limits on
$N($\ion{H}{1}$)$ based on changes in \kid, we then increased (or
decreased) the \ion{H}{1} column density while allowing the other
parameters to vary. By allowing the polynomial coefficients to vary
throughout the entire process, the fitting procedure can adjust the
height and curvature of the continuum and thereby provide some
assessment of the continuum placement uncertainty.  With {\it IUE}
data, background subtraction near
\lya\ is difficult due to order crowding, and this introduces
significant uncertainty in the flux zero point.  For this reason, it
is important to include the flux zero point as a free parameter in the
fitting process. Our final best fit and the corresponding continuum
placement are shown in Fig.~\ref{fig_HI_iue} along with the profiles
corresponding to \1s\ upper and lower limits on $N($\ion{H}{1}$)$.
From these fits we obtain $\log N($\ion{H}{1}$) = 21.17 \pm0.10$ (\2s\
uncertainties).

\begin{figure}
\begin{center}
\psfig{file=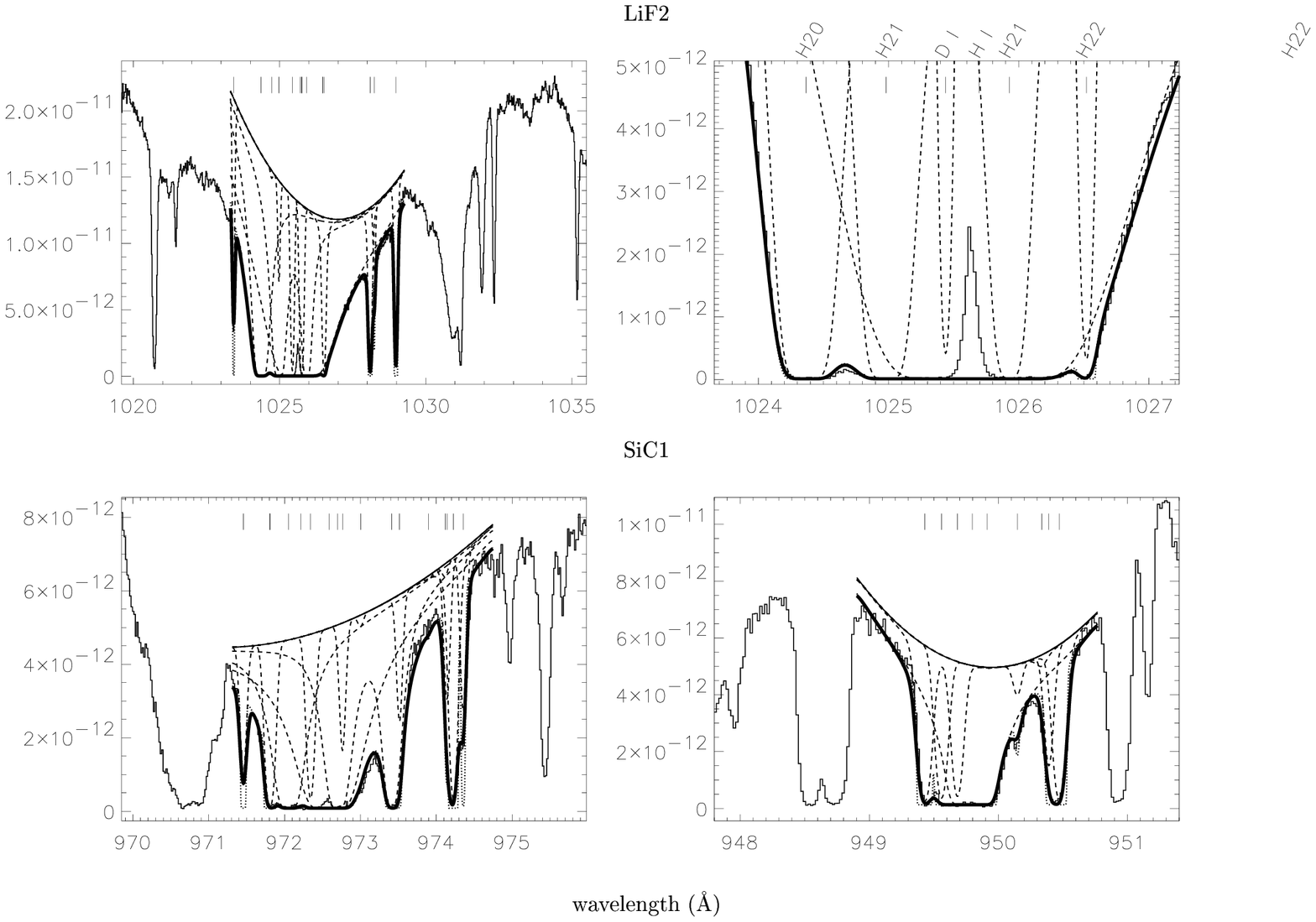,height=15cm}
\caption{Fit of the \lyb, \lyg, and \lyd\ \ion{H}{1} lines observed 
with \fuse\ toward \hd9.  Same conventions as in Fig.~\ref{fig_fit}
but for clarity, the species are not identified at the top of the
plots on three of the four panels.  The emission lines in the cores of
the \ion{H}{1} absorptions (near 1025.65\,\AA\ and 972.55\,\AA) are
geocoronal.
The result is in agreement with the measurement from \lya\ ({\it IUE}), but
the stellar continua are difficult to constrain, especially for~\lyb.
\label{fig_HI_fuse}
}
\end{center}
\end{figure}

This result is in good agreement with the values obtained from the
fits of \lyb, \lyg, and \lyd\ observed with \fuse. We obtained $\log
N($\ion{H}{1}$)=21.19\pm0.10$ from the fits of these three lines
(Fig.~\ref{fig_HI_fuse}), using all the available data. However, if
\lyb\ presents damping wings, the stellar continuum at this wavelength
is a critical issue. It is probably reached by strong \ion{O}{6}
features due to the stellar wind (see, e.g, Bianchi \&
Garcia~\citealp{bianchi02}). No robust models of the stellar
continuum are available up to now. The polynomials that we used give
an estimation of the stellar continua shape, but they are probably too
naive. Thus, we adopt the \lya\ measurement for our $N($\ion{H}{1}$)$
result.

On this line of sight, a minority of the material is in molecular
form.  The hydrogen molecular fraction $f($H$_2) \equiv 2N($H$_2) / [2
N($H$_2) + N($\ion{H}{1}$)] = 10.1\pm2.6$~\%.  
Thus, molecular species can be neglected when the abundances are
computed from column densities ratios. The implied errors on abundance
would be of the order of 10~\%,
probably even less, as the molecular species are unlikely to
significantly coexist in the same clouds with the atomic and ionic
species. By comparison with the final reported error bars, the effect
of the molecules on the abundance is negligible. For example, we 
obtain \ion{D}{1}/\ion{H}{1}$\,=(9.8\pm3.8)\times10^{-6}$ and 
(\ion{D}{1}+HD)/(\ion{H}{1}+2H$_2)=(8.9\pm3.3)\times10^{-6}$.

On the contrary to H$_2$ and HD (\S~\ref{HD}), we did not detect any
significant velocity shifts between the \ion{D}{1} and \ion{H}{1}
lines ($< 1$~\kms) on the \fuse\ spectra, which is reassuring as the
ratio of the column densities of these two species are directly used
to compute the D/H ratio. Some targets have been reported with
significant shifts between the \ion{D}{1} and \ion{H}{1} lines. Wood
et al.~(\citealp{wood04}), for example, reported a shift of 5~\kms\
between the radial velocities of \ion{H}{1} and \ion{D}{1} on the
sight line of JL\,9.

\subsection{Discussion}

The abundances on the line of sight to \hd9\ are reported in
Table~\ref{table_ratios}.  The D/H, D/O, and D/N ratios are all
significantly lower than the ratios measured within the Local
Bubble. This is in agreement with the picture reported by H\'ebrard \&
Moos~(\citealp{hebrard03}), which suggests that the deuterium
abundance is significantly lower in the distant interstellar medium
than~locally.

The N/H ratio is in good agreement with the typical interstellar N/H
ratio reported by Meyer et al.~(\citealp{meyer97}). The O/H ratio is
rather large when compared with the typical O/H ratios in the distant
interstellar medium (Meyer et al.~(\citealp{meyer98}; Andr\'e et
al.~\citealp{andre03}; Cartledge et al.~\citealp{cartledge04}),
although compatible within the error bars.  Other targets have been
found to show similar O/H ratios: HD\,91824, HD\,122879, HD\,12323
(Cartledge et al.~\citealp{cartledge04}), HD\,195965 (Hoopes et
al.~~\citealp{hoopes03}), and JL\,9 (Wood et al.~\citealp{wood04}). It
can be argued that the low D/O observed toward \hd9\ is due to the
high O/H ratio rather than to a true low deuterium abundance. However,
the D/H and D/N ratios are low as well. In addition, if we assume the
\ion{O}{1} column density is overestimated by a factor $\sim1.6$ (in
order to agree with the above O/H studies), this would still imply a
low D/O, below $3\times10^{-2}$.  The deuterium abundance seems thus
to be actually low on this line of sight. It is interesting to note
that \hd9\ shows simultaneously a low deuterium abundance together
with a rather high oxygen abundance. HD\,195965 (Hoopes et
al.~\citealp{hoopes03}) and JL\,9 (Wood et al.~\citealp{wood04}) also
show a similar effect. This might be a hint of an astration signature.

The HD/2H$_2$ ratio is about 5 times lower than the D/H ratio.  This
is characteristic of interstellar clouds with moderate molecular
fraction (We obtain $f($H$_2) = 10.1\pm2.6$~\%).  
Indeed, compared to H$_2$, self-shielding of HD becomes significant
deeper in the molecular clouds since deuterium is much less abundant
than hydrogen (Ferlet et al.~\citealp{ferlet00}; Lacour et
al.~\citealp{lacour05}). Thus, the transition between \ion{D}{1} and
HD takes place deeper in the molecular clouds than the transition
between \ion{H}{1} and H$_2$. Our HD/2H$_2$ ratio is similar to those
reported by Lacour et al.~(\citealp{lacour05}).

\section{The \feige\ line of sight}
\label{feige}

\subsection{The target}

An extensive study of the \feige\ sight line was performed by Friedman
et al.~(\citealp{friedman02}). All the information about this target
and its observations are in that initial paper.  Briefly, \feige\ is a
sdOB star.
Most of the interstellar material probed along its sight line is
probably located outside the Local Bubble. The parallax and
photometric distances, $179^{+265}_{-67}$~pc and $288 \pm 43$~pc
respectively, are in good agreement.  Its high Galactic latitude
(-59\fdg07) locates \feige\ at $z\simeq-150$~pc from the Galactic
plane.

The \fuse\ spectra of \feige\ were obtained in 2000
(Table~\ref{table_obslog}). Friedman et al.~(\citealp{friedman02})
measured $N($\ion{D}{1}$)$ and $N($\ion{O}{1}$)$ from these spectra,
and used {\it IUE} archival data to derive $N($\ion{H}{1}$)$ using the
same technique we applied to \hd9\ (\S\ref{HI}). Subsequently,
H\'ebrard \& Moos~(\citealp{hebrard03}) used the \fuse\ spectra of
\feige\ to measure $N($\ion{N}{1}$)$ on this sight line.

The \ion{O}{1} column density measurement reported by Friedman et
al.~(\citealp{friedman02}) was derived using only a curve-of-growth
analysis. However, the weak \ion{O}{1} line at $\lambda\,974.07\,$\AA\
(see \S~\ref{hd900987_OI}) was not included in this early analysis,
and all the \ion{O}{1} lines fitted in the curve-of-growth are likely
to be significantly saturated. As saturated lines might lead to
erroneous column density evaluations (Friedman et
al.~\citealp{friedman05}), we revisit here the $N($\ion{O}{1}$)$
measurement toward~\feige.

\subsection{The \ion{O}{1} column density revisited}

We used exactly the same dataset previously employed by Friedman et
al.~(\citealp{friedman02}) and H\'ebrard \&
Moos~(\citealp{hebrard03}). As for \hd9\ (\S\ref{Overview}), we
measured the column densities on the sight line toward \feige\ using
the profile fitting method presented by H\'ebrard et
al.~(\citealp{hebrard02}), which is based on the procedure \ow\ 
(Lemoine et al.~\citealp{lemoine02}). \ion{D}{1}, \ion{O}{1},
\ion{N}{1}, and H$_2$ lines were included in the fits. We focus here 
on \ion{O}{1}. We note that we obtain results in agreement with
Friedman et al.~(\citealp{friedman02}) and H\'ebrard \&
Moos~(\citealp{hebrard03}) for $N($\ion{D}{1}$)$ and 
$N($\ion{N}{1}$)$,~respectively.

\begin{figure}
\begin{center}
\psfig{file=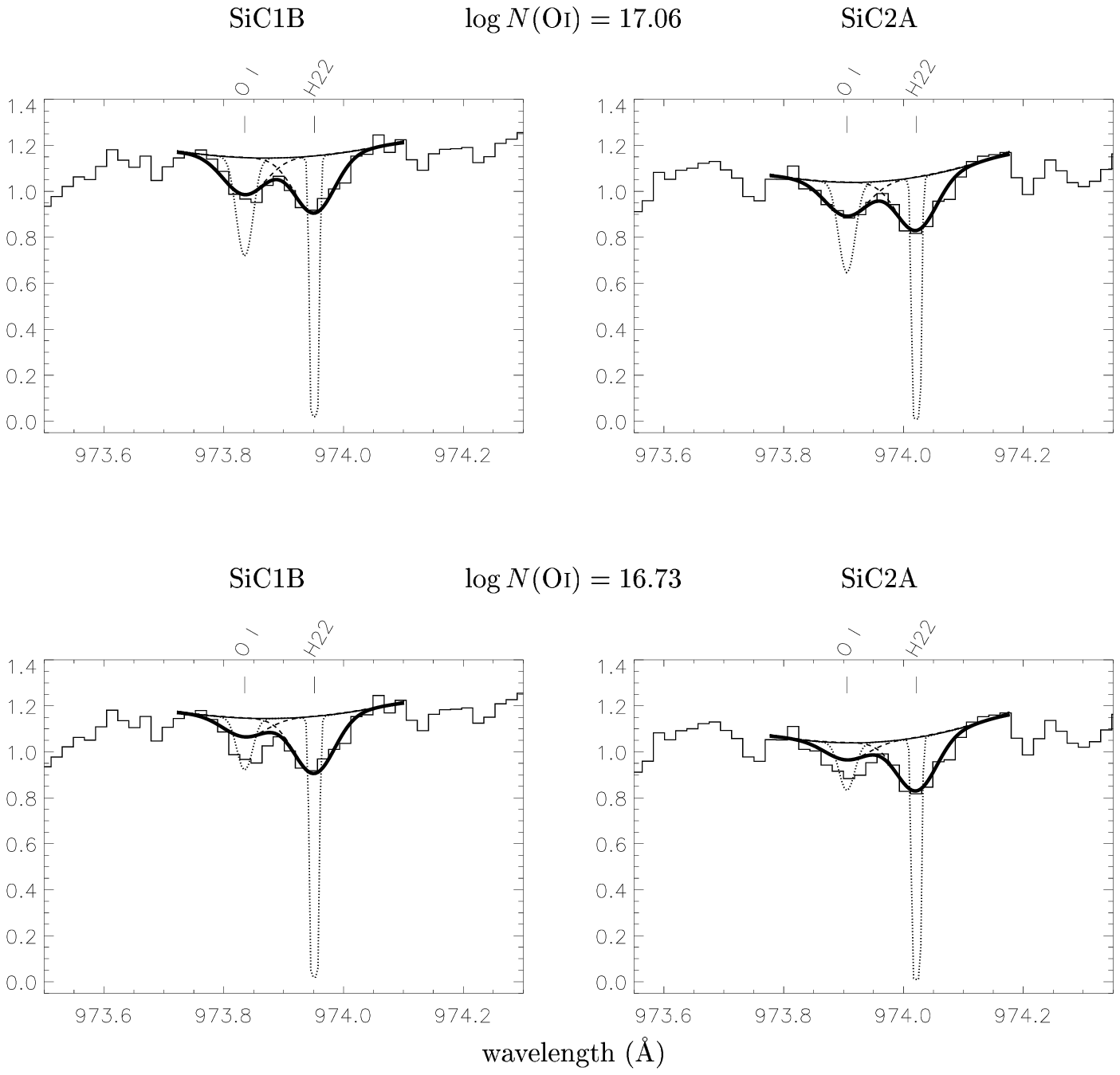,height=12cm}
\caption{Fits of the $\lambda\,974.07\,$\AA\ \ion{O}{1} line toward 
\feige. Same conventions as in Fig.~\ref{fig_fit}. The Y-axis is flux in 
$10^{-11}$\,erg/cm$^2$/s/\AA. The upper panel shows our fit of the
SiC1B and SiC2A data. The lower panel shows the fit of the same data
assuming the \ion{O}{1} column density reported by Friedman et
al.~(\citealp{friedman02}). The $\lambda\,974.07\,$\AA\ line suggests
this early $N($\ion{O}{1}$)$ measurement was underestimated by a
factor $\sim2$.
\label{fig_OI_feige}
}
\end{center}
\end{figure}

As for \hd9, we found that $\lambda\,974.07\,$\AA\ is the only
unsaturated \ion{O}{1} line available in the \fuse\ bandpass. However,
this line is less blended in the case of \feige:
$\lambda\,974.07\,$\AA\ is blended with only one H$_2$ ($J=2$)
absorption (see Fig.~\ref{fig_OI_feige}). The shape of this saturated
H$_2$ line is constrained thanks to the other H$_2$ transitions
available in the \feige\ \fuse\ spectra. The two other H$_2$
transitions located around 974.07\,\AA\ (corresponding to the level
$J=5$ and $J=6$) are negligible for this sight line. Our fits include
the two different available observations of $\lambda\,974\,$\AA\
(segments SiC1B and SiC2A).

\begin{figure}
\begin{center}
\psfig{file=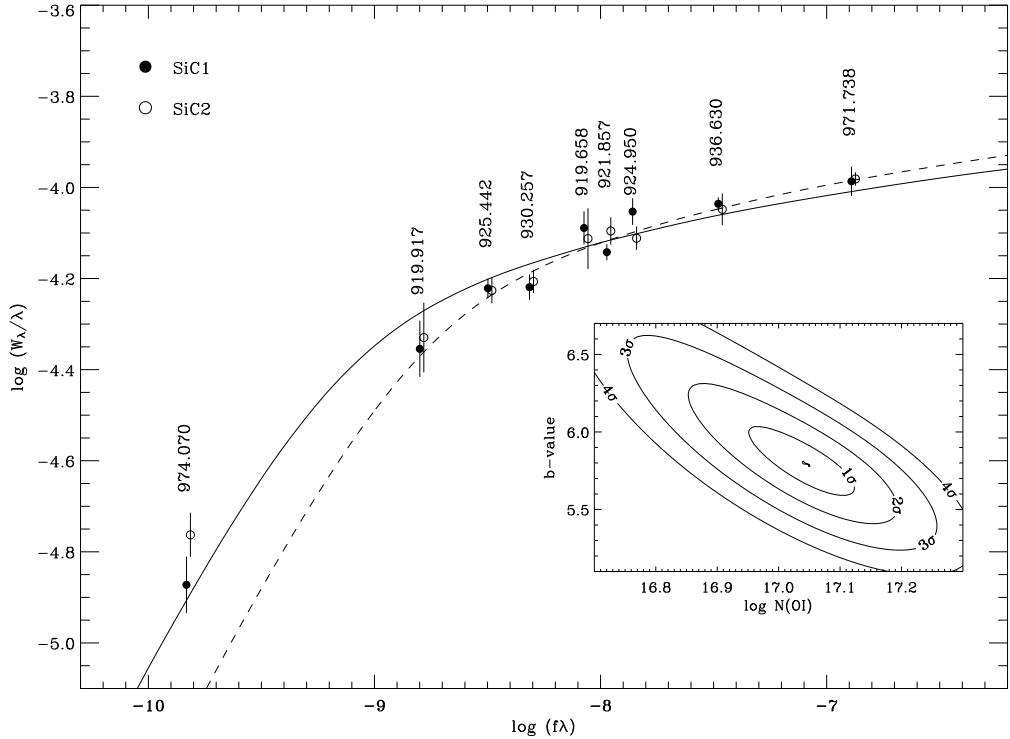,height=14cm,angle=90}
\caption{The single component, Gaussian curve-of-growth for \ion{O}{1} 
toward \feige.  The best fit solution (solid line) is log
$N($\ion{O}{1}$) = 17.04^{+0.15}_{-0.18}$ and $b =
5.80^{+0.51}_{-0.39}$. The solution if the $\lambda$\,974\,\AA\ line
is not included (dashed line) is $16.73\pm{0.10}$ and $b =
6.58\pm{0.35}.$ The inset show the log($N$)/$b-$value error
contours. For clarity, at each wavelength the data points derived from
each of the \fuse\ channels have been slightly separated.  Including
saturated lines might lead to erroneous column density measurements
and underestimations of error~bars.
\label{fig_OI_COG_feige}
}
\end{center}
\end{figure}

We checked the results of our fits by performing a curve-of-growth
analysis, using the same procedure as Friedman et
al.~(\citealp{friedman02}). We normalized the spectrum by a fitted
continuum and a synthetic H$_2$ model constrained by the other H$_2$
transitions. Then, as we did for the other, stronger \ion{O}{1} lines,
we measured the equivalent width of $\lambda\,974\,$\AA. We obtained
$13.1\pm2.7$\,m\AA\ and $16.8\pm2.7$\,m\AA\ for the segments SiC1B and
SiC2A, respectively.  The resulting curve of growth is plotted in
Fig.~\ref{fig_OI_COG_feige}.  The column densities measured using
these two methods (profile fitting and curve of growth) are different
by only 0.03\,dex. This difference is negligible with respect to the
uncertainties ($\pm 0.15$\,dex, \2s), so there are apparently no
systematic effects due to the method.

Our final value, which reflects the combined effort of these two
independent analyses, is $\log N($\ion{O}{1}$) = 17.06 \pm0.15$. This
is a factor $\sim2$ larger than the one reported by Friedman et
al.~(\citealp{friedman02}), namely $\log N($\ion{O}{1}$) = 16.73
\pm0.10$ (\2s). Fig.~\ref{fig_OI_feige} shows the fit of 
$\lambda\,974\,$\AA\ using these two $N($\ion{O}{1}$)$ values, which
clearly suggests a higher value than the one reported by Friedman et
al. This illustrates that measuring column densities from saturated
lines might lead to erroneous results and underestimation of
error~bars.

Our past experience with different \fuse\ targets together with
simulations of \fuse\ data (Moos et al.~\citealp{moos02}; H\'ebrard et
al.~\citealp{hebrard02}; Friedman et al.~\citealp{friedman05}) has
shown that measurements from saturated lines seem to preferentially
underestimate $N($\ion{O}{1}$)$. The case of \feige\ is in agreement
with this tendency. We note that the value of $f\lambda$ for
$\lambda\,974\,$\AA\ is at least 10 times smaller than for the other
\ion{O}{1} lines in the \fuse\ bandpass. Such large differences might 
explain significant systematic effects due to saturation in the range
$\log N($\ion{O}{1}$)\simeq17-18$ when the $\lambda\,974\,$\AA\
transition is not used in the analysis. In addition, one can note that
the equivalent width of the $\lambda\,919.917\,$\AA\ \ion{O}{1} line
seems to indicate a lower column density (see
Fig.~\ref{fig_OI_COG_feige}).  The reason is unclear, but it can be due to
structure along the sight line, some components being saturated before
the other ones. A slight inaccuracy in $f$-value for the
$\lambda\,919.917\,$\AA\ transition and/or blending issue implying
continuum uncertainties in that spectral area might also be present.
Such problems with $\lambda\,919.917\,$\AA\ might increase the error
for $N($\ion{O}{1}$)$ even more, if $\lambda\,974\,$\AA\ is not used.

Targets in that column density range ($\log
N($\ion{O}{1}$)\simeq17-18$) and with poor signal to noise should be
considered with caution. Indeed, if the signal to noise is not high
enough to allow $\lambda\,974\,$\AA\ to be detected, $N($\ion{O}{1}$)$
measurements might be significantly underestimated (see the dashed
line in Fig.~\ref{fig_OI_COG_feige}).

Since it is mainly based on one transition, our final result on
$N$(\ion{O}{1}) toward \feige\ is highly sensitive to possible errors
caused by uncontrolled blends from unknown lines (numerous stellar
lines due to metals are detected in the spectrum of \feige) or an
erroneous $f$-value. Up to now, studies of other targets have not
suggested such problems (see \S~\ref{hd900987_OI} and Friedman et
al.~\citealp{friedman05}).

\subsection{Discussion}

Using our new \ion{O}{1} column density measurement together with the
$N($\ion{D}{1}$)$ reported by Friedman et al.~(\citealp{friedman02}),
we obtain D/O~$=(2.6\pm1.0) \times 10^{-2}$ (\2s).  This ratio is
lower than D/O measured in the Local Bubble, \dsolb$=(3.84\pm0.16)
\times 10^{-2}$ (\1s, H\'ebrard \& Moos~\citealp{hebrard03}). 
The lower D/O value is in agreement with the fact that the deuterium
abundance might be lower for the most distant lines of sight and the
highest column densities, as reported by H\'ebrard \&
Moos~(\citealp{hebrard03}).

Friedman et al.~(\citealp{friedman02}), however, reported
D/H~$=(2.14\pm0.82) \times 10^{-5}$ (\2s) on this sight line. This
ratio is {\it larger} than the Local Bubble value, which is in the
range $(1.3-1.5) \times 10^{-5}$ (see \S~\ref{Introduction}). This
ratio is also about three times larger than the D/H typical from the
distant interstellar medium reported by H\'ebrard \&
Moos~(\citealp{hebrard03}). The apparent contradiction between D/H and
D/O is surprising. Thus, this sight line appears atypical.

According to our new evaluation of $N($\ion{O}{1}$)$ and the
$N($\ion{H}{1}$)$ from Friedman et al.~(\citealp{friedman02}), the
oxygen abundance toward \feige\ is O/H~$=(8.3\pm4.6) \times 10^{-4}$
(\2s). This ratio is about two times larger than the typical O/H
ratios reported by Meyer et al.~(\citealp{meyer98}), Andr\'e et
al.~(\citealp{andre03}), or Cartledge et al.~(\citealp{cartledge04}).

Similarly, the $N($\ion{N}{1}$)$ measurement from H\'ebrard \&
Moos~(\citealp{hebrard03}) and the $N($\ion{H}{1}$)$ measurement from
Friedman et al.~(\citealp{friedman02}) imply N/H~$=(2.4\pm1.5) \times
10^{-4}$ (\2s) for \feige. This ratio is about three times larger than
the typical N/H ratios reported by Meyer et~al.~(\citealp{meyer97}). 

The fact that D/H, O/H, N/H are all approximately two-three times
larger than the values usually measured in the distant interstellar
medium is suspicious. It suggests that the \ion{H}{1} column density
might be affected by some systematic effects, and the actual
$N($\ion{H}{1}$)$ being two-three times larger.  However, the IUE
\lya\ spectrum of \feige\ studied by Friedman et
al.~(\citealp{friedman02}) appears not consistent with such a high
\ion{H}{1} column density. New \lya\ observations would be very useful 
for that target but due to the recent failure of the spectrograph STIS
onboard HST, such observations won't be possible at short term.
Whatever the solution is, \feige\ appears as an abnormal line of
sight, with abnormally high D/H, O/H, and N/H~ratios.

Finally, the Ti/H ratio appears also particularly high toward
\feige\ (Prochaska et al.~\citealp{prochaska05}). This point is important, 
as \feige\ is the target that mainly drives the correlation between
\ion{D}{1} and \ion{Ti}{2} recently reported by Prochaska et
al.~(\citealp{prochaska05}). Such a correlation might be interpreted
as a support for a significant depletion of deuterium onto dust grains
(Linsky et al.~\citealp{linsky05}), as Ti is a refractory
element. However, the systematic effect that we suspect on
$N($\ion{H}{1}$)$ might also be the cause for the high Ti/H ratio
measured toward \feige.

\section{Comparison of D/H and D/O measurements}

In this section we compare our new D/H and D/O measurements with those
obtained toward other lines of sight.  D/O is considered to be a good
proxy for D/H and, in addition, it may be less sensitive to systematic
effects (Timmes et al.~\citealp{timmes97}; H\'ebrard \&
Moos~\citealp{hebrard03}).

\begin{figure}
\begin{center}
\psfig{file=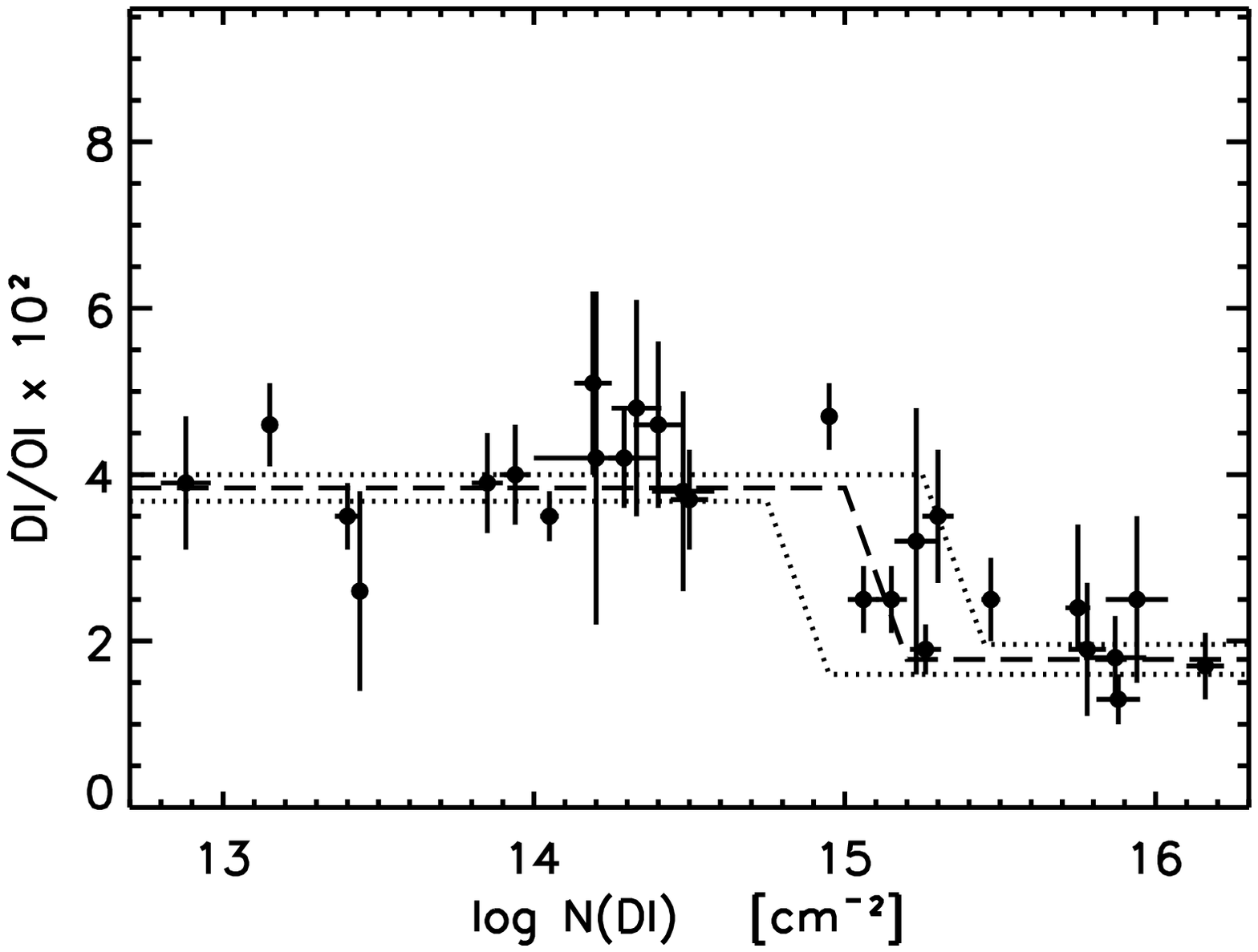,height=5.8cm}
\psfig{file=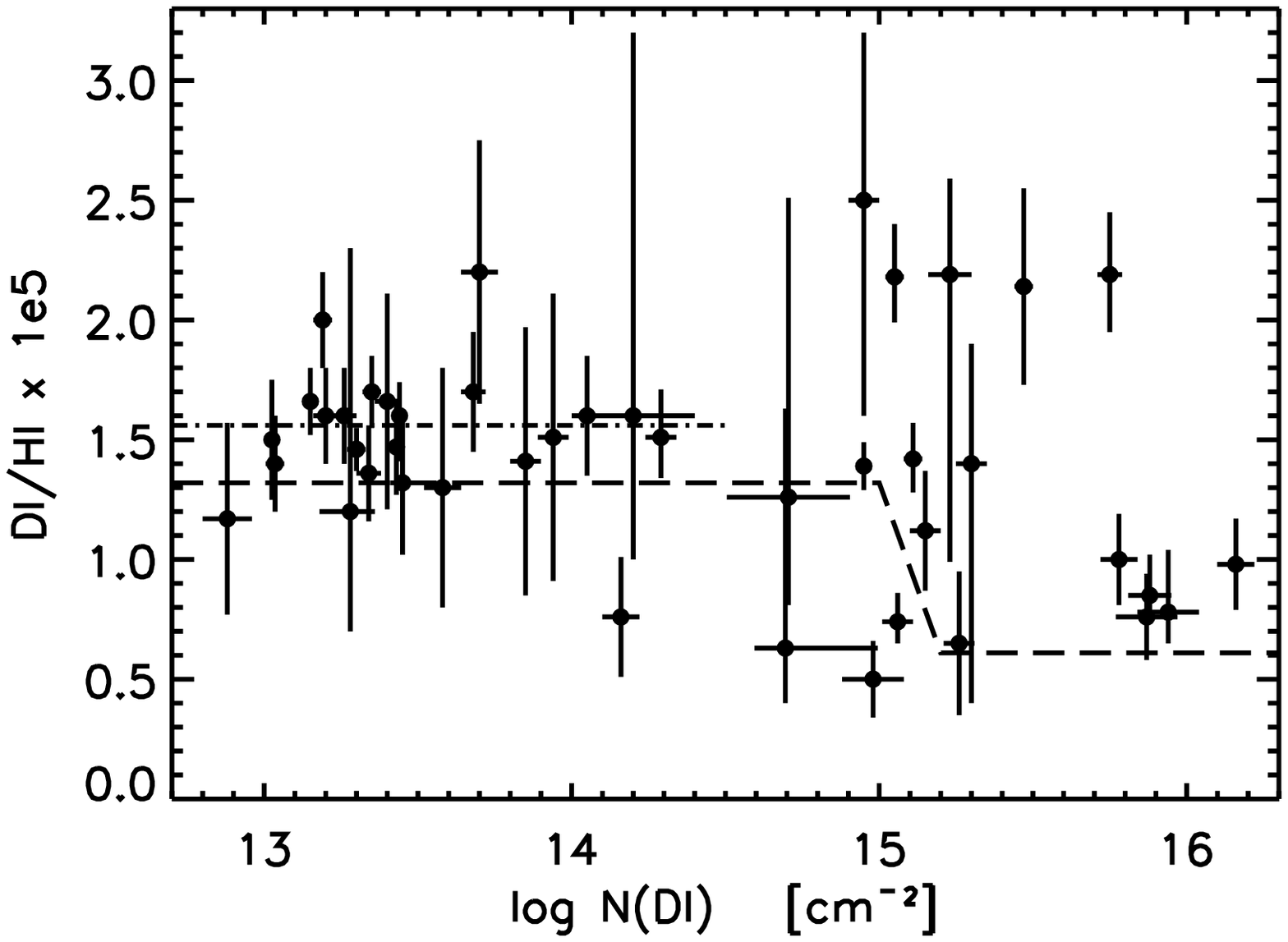,height=5.8cm}
\vspace{0.4cm}
\caption{D/O (left) and D/H (right) as a function of $\log N($\ion{D}{1}$)$. 
The $y$-axes of the two plots are scaled to the same size, assuming
O/H$\;=3.43\times 10^{-4}$ (Meyer~\citealp{meyer01}). The D/O plot
(left) shows a simple, bimodal picture (dashed line, with dotted error
bars): there are a local and a distant D/O ratio, respectively
$(3.84\pm0.16) \times 10^{-2}$ and $(1.78\pm0.18) \times 10^{-2}$,
with a transition around $\log N($\ion{D}{1}$)=15-15.3$.  The
extension of the transition area is not well constrained with the
currently available dataset. The D/H plot (right) shows a less simple,
more scattered picture, which doesn't fit with the D/O bimodal picture
(dashed line, over-plotted here for comparison). The D/O ratios
measured on distant sight lines ($\log N($\ion{D}{1}$)>15$) are all
low, whereas there are targets in the same column density range which
present high D/H ratios. Note that the local D/H ratio ($1.56\pm0.04)
\times10^{-5}$ (Wood et al.~\citealp{wood04}, dashed-dotted line) is
significantly higher the local D/H ratio inferred from D/O
measurements (see H\'ebrard \& Moos~\citealp{hebrard03}). All the
error bars plotted here are \1s.
\label{fig_dso_dsh}
}
\end{center}
\end{figure}

Fig.~\ref{fig_dso_dsh} (left panel) shows D/O as a function of the
\ion{D}{1} column density. This plot includes 27
sight lines: the 24 included in H\'ebrard \&
Moos~(\citealp{hebrard03}), but with the revised $N$(\ion{O}{1}) for
\feige\ (\S~\ref{feige}), plus JL\,9 (Wood et al.~\citealp{wood04}), 
PG\,0038+199 (Williger et al.~\citealp{williger05}), and HD\,90087
(\S~\ref{hd90087}). Thus, D/O measurements apparently show a bimodal
picture: there is a high, homogeneous D/O ratio for low \ion{D}{1}
column densities, and a low, homogeneous D/O ratio for high \ion{D}{1}
column densities. The weighted mean for the 14 lines of sight within
the Local Bubble is \dsolb$\,= (3.84\pm0.16) \times 10^{-2}$
(H\'ebrard \& Moos~\citealp{hebrard03}).  The weighted mean for the 7
lines of sight with the highest $N$(\ion{D}{1}) is about two times
lower: D/O$\,= (1.75\pm0.18) \times 10^{-2}$, with a \kid\ of 5.9 for
6 degrees of freedom (reduced \kid\ of 1.0). We do not see any
variations in the distant D/O ratio. These 7 sight lines are
Feige\,110 (Friedman et al.~\citealp{friedman02} and
\S~\ref{feige}), HD\,195965 and HD\,191877 (Hoopes et
al.~\citealp{hoopes03}), LSS\,1274 (H\'ebrard \&
Moos~\citealp{hebrard03}; Wood et al.~\citealp{wood04}), JL\,9 (Wood
et al.~\citealp{wood04}), PG\,0038+199 (Williger et
al.~\citealp{williger05}), and HD\,90087 (\S~\ref{hd90087});
they all present $\log N($\ion{D}{1}$)>15.4$. The transition between
the two D/O ratios, $3.84 \times 10^{-2}$ and $1.75 \times 10^{-2}$,
is located around $\log N($\ion{D}{1}$)\simeq15-15.3$.

This simple, bimodal picture of D/O is rather different from what D/H
shows. Indeed, a larger dispersion is seen through D/H measurements.
On Fig.~\ref{fig_dso_dsh} (right panel) are plotted D/H ratios as a
function of $\log N($\ion{D}{1}$)$ for 
43 lines of sight. This includes the 39 targets reported in the
Table~4 from Wood et al.~(\citealp{wood04}), to which we added Sirius
(H\'ebrard et al.~\citealp{hebrard99}), Lan\,23 (Oliveira et
al.~\citealp{oliveira03}), PG\,0038+199 (Williger et
al.~\citealp{williger05}), and HD\,90087 (\S~\ref{hd90087}).
Note that we adopt D/H~$=(2.2^{+0.4}_{-1.2})\times10^{-5}$ for
Lan\,23, according the \ion{D}{1} and \ion{H}{1} column densities
reported by Oliveira et al.~(\citealp{oliveira03}). The
$N($\ion{H}{1}$)$ toward this target is derived from low EUVE flux,
which makes the measurement uncertain, especially for such high column
density. The $x$-axes of the two plots of Fig.~\ref{fig_dso_dsh} are
identical, and their $y$-axes are scaled to the same size (assuming
O/H$\;=3.43\times 10^{-4}$; Meyer~\citealp{meyer01}); this allows the
two panels to be easily compared.

Within the Local Bubble, at low \ion{D}{1} column densities, both
panels show a homogeneous deuterium abundance (note the slight but
significant disagreement between the two averaged values; see
H\'ebrard \& Moos~\citealp{hebrard03}).  However, the two panels
present different pictures at larger column densities: whereas D/O
shows a decrease toward a homogeneous value, D/H shows dispersions.
Considering as above the same 7 lines of sight with $\log
N($\ion{D}{1}$)>15.4$, the \kid\ for the weighted mean of this D/H
distribution is 34.4, whereas it is 5.9 in the case of D/O (for 6
degrees of freedom in both cases). D/H seems more dispersed than D/O
in the distant interstellar medium.

Since oxygen is considered as a good proxy for hydrogen, these two
different pictures are unexpected.  This discrepancy might call into
question the homogeneity of the O/H ratio within the interstellar
medium, but numerous studies show that \ion{O}{1} is a good tracer of
\ion{H}{1} in the nearby Galactic disk (Meyer et
al.~\citealp{meyer98}; Andr\'e et al.~\citealp{andre03}; Cartledge et
al.~\citealp{cartledge04}; Oliveira et al.~\citealp{oliveira05}).
Andr\'e et al.~(\citealp{andre03}) and Cartledge et
al.~(\citealp{cartledge04}) suggested however O/H can vary somewhat as
a function of distance and/or average density. On another hand, it can
be argued that some \ion{H}{1} column densities are under-evaluated,
as it may be possible for \feige\ (\S~\ref{feige}), but no systematics
allowing such under-evaluation have been yet clearly identified.
Fig.~13 in Williger et al.~(\citealp{williger05}) shows explicitly how
poor the PG\,0038+199 \lyb\ line is fitted if $N($\ion{H}{1}$)$ is
forced to be high enough to recover the expected D/H and O/H values.
\ion{O}{1} column densities might also be suspected to be erroneous for 
distant lines of sight, as it is mainly measured from only one
transition, namely $\lambda\,974.07\,$\AA. This transition appears
however to be reliable (\S~\ref{hd900987_OI}), and there are several
targets that show low D/O ratios whereas their $N($\ion{O}{1}$)$
measurements are based on $\lambda\,1356\,$\AA\ ($\delta\,$Ori$\,$A,
$\gamma\,$Cas, $\epsilon\,$Ori, HD\,195965; see H\'ebrard \&
Moos~\citealp{hebrard03}).  Finally, if deuterium is significantly
depleted onto dust grains, as proposed by Wood et
al.~(\citealp{wood04}) and Linsky et al.~(\citealp{linsky05}), one
would have to argue that oxygen is depleted at the same rate, in order
explain the observed constancy of D/O and variability of D/H. Such a
tight correlation between deuterium and oxygen depletion seems
unlikely, however. Indeed, oxygen tends to be associated with
silicates, whereas Draine~(\citealp{draine04}) proposed that PAHs are
causing the deuterium depletion.  Up to now, the discrepancy between
the constancy of D/O and the variability of D/H remains
unexplained. Several processes might be implied, including the above
ones, but the complete explanation is still to be found.

It should also be recalled that distant targets with deuterium
measurements remain sparse.  Actually, only three lines of sight
exhibit significantly high D/H ratios, namely PG\,0038+199,
Feige\,110, and $\gamma^2$\,Vel.  These targets are important as they
are those that are assumed to be characteristic of the actual \dshpe\ 
value if deuterium is significantly depleted onto dust grains.  For
the two first ones, both D/H and D/O measurements are available.  For
PG\,0038+199, D/H$\,= (2.19^{+0.30}_{-0.27}) \times 10^{-5}$ and
D/O$\,= (2.40^{+1.03}_{-0.45}) \times 10^{-2}$ (Williger et
al.~\citealp{williger05}), and for Feige\,110, D/H$\,= (2.14\pm0.41)
\times 10^{-5}$ (Friedman et al.~\citealp{friedman02}) and 
D/O$\,= (2.6\pm0.5) \times 10^{-2}$ (\S~\ref{feige}).
Thus, both present ``low'' D/O ratios, but ``high'' D/H.  No direct
$\log N($\ion{O}{1}$)$ measurements are available for the last target,
$\gamma^2$\,Vel, for which Sonneborn et al.~(\citealp{sonneborn00})
reported D/H$=(2.18^{+0.22}_{-0.19}) \times 10^{-5}$.  The only
available estimation is reported by Knauth et
al.~(\citealp{knauth03}):
$N($\ion{O}{1}$)=(5.23\pm1.27)\times10^{16}$\cmmd. This is based on 
\ion{O}{1} and \ion{S}{2} observations by Fitzpatrick \& 
Spitzer~(\citealp{fitzpatrick94}).  It leads to a low D/O ratio ($\sim
2.1 \times 10^{-2}$), but the systematic uncertainties in this
$N($\ion{O}{1}$)$ indirect estimation are too large to allow any firm
conclusion to be drawn. Indeed, using updated solar abundances and a
lower depletion factor might produce D/O up to $\sim3.5$ times larger.

More D/H and D/O measurements toward distant lines of sight are
mandatory to compare the enigmatic behavior of these ratios.

\section{Conclusion}

We have presented a new interstellar deuterium abundance study of the
sight line toward \hd9, and the revision of a result for \feige, both
based on \fuse\ spectra.

Our new measurement of $N($\ion{O}{1}$)$ toward \feige\ shows that
\ion{O}{1} column density measurements based on saturated lines might
be erroneous, even up to a factor of two, and that the error bars
might be underestimated due to systematic effects.  This sight line
exhibits a low D/O ratio,~in~agreement with those obtained toward
other distant~targets. 

\hd9\ is the farthest Galactic target for which the deuterium abundance 
has been measured from ultraviolet absorption lines so far, both in
terms of distance and column densities. The low deuterium abundance
measured on this line of sight is in agreement with the trend reported
by H\'ebrard \& Moos~(\citealp{hebrard03}) and reinforced by Wood et
al.~(\citealp{wood04}): on large scales, the deuterium abundance is
lower than the one measured within the Local Bubble.  The question
however remains open to know if this low deuterium abundance is the
value representative of the present epoch, or the signature that a
significant amount of deuterium has been depleted onto dust grains.

Finally, we have shown that D/H presents more dispersion than D/O
does.  As oxygen is considered to be a good proxy for hydrogen, this
discrepancy is puzzling.  This might call into question the
homogeneity of the O/H ratio, or some \ion{H}{1} or \ion{O}{1} column
density measurements. The complete explanation is still to be found.

\acknowledgments
This work is based on data obtained by the NASA-CNES-CSA \fuse\ 
mission operated by the Johns Hopkins University. Financial support
has been provided by NASA contract NAS5-32985.  We thank D. Massa and
A. W. Fullerton for their help on the stellar profile of \hd9, as well
as C. M. Oliveira and A. Vidal-Madjar for useful comments on this
paper. Most of this work was performed during the stay of G.\,H. at
the Johns Hopkins University in 2004.  G.\,H. was also supported by
CNES.  T.\,M.\,T. was supported in part by NASA grant
NNG\,04GG73G. This work used the profile fitting procedure Owens.f
developed by M. Lemoine and the French \fuse\ Team.

%%%%%%%%%%%%%%%%%%%%%%%%%%%%%%%%%%%%%%%%%%%%%%%%%%%%%%%%%%%%%%%%%%%

%%%%%%%%%%%%%%%%%%%%%%%%%%%%%%%%%%%%%%%%%%%%%%%%%%%%%%%%%%%%%%%%%%%
%\clearpage

%\clearpage

%\clearpage

%\clearpage

%\clearpage

%\clearpage

%\clearpage

%\clearpage

%\clearpage

%\clearpage

%\clearpage

%\clearpage

%\clearpage

%\clearpage

%\clearpage
%%%%%%%%%%%%%%%%%%%%%%%%%%%%%%%%%%%%%%%%%%%%%%%%%%%%%%%%%%%%%%%%%%%

\begin{table}
\begin{center}
\caption{\hd9\ summary.} 
\begin{tabular}{lcc}
\hline
\hline
Quantity & Value & Reference \\
\hline
Spectral Type               & O9.5 III                   & 1 \\
Right ascension (2000)      & 10:22:20.8                 & 2 \\
Declination (2000)          & $-$59:45:19.7   & 2 \\
Galactic longitude ($l$)    & $285.15^{\circ}$           & 2 \\
Galactic latitude ($b$)     & $-2.13^{\circ}$            & 2 \\
Distance ($d$)              & $2740$~pc                  & 3 \\
Distance to Galactic plane ($z$) & $-102$~pc       & 3 \\
Magnitude ($V$)             & $7.80$                     & 3 \\
$E(B-V)$                    & 0.28                       & 3 \\
Rotational velocity ($v \sin i$) & 272~\kms              & 3 \\
Radial velocity ($V_{\rm{rad}}$)  & $-2$~\kms\           & 3 \\
\hline
\end{tabular}
\label{table_star_parameters}
\\
\vspace{0.1cm}
References -- 
(1) Mathys~\citealp{mathys88}; 
(2) Hog et al.~\citealp{hog98};
(3) Savage, Meade, \& Sembach~\citealp{savage01}.
\end{center}
\end{table}

%\clearpage

\begin{table*}
\begin{center}
\caption{Log of the observations.} 
\begin{tabular}{ccccrrc}
\hline
\hline
Target & Instrument &
Obs. date & data ID & $T_{\rm obs}^{\rm a}$ & $N_{\rm exp}^{\rm b}$ &
Apert.$^{\rm c}$ \\
\hline
\hd9   & \fuse & 2000 Apr 03 & P1022901 &  3.9 &  5 & LWRS \\
\hd9   & \fuse & 2003 Jun 17 & P3030401 & 14.0 & 32 & MDRS \\
\hd9   & {\it IUE}   & 1982 Mar 14 & SWP16532 &  3.9 &  1 & LARGE \\
\hline
\feige & \fuse & 2000 Jun 22 & M1080801 &  6.2 &  8 & LWRS \\
\feige & \fuse & 2000 Jun 30 & P1044301 & 21.8 & 48 & LWRS \\
\hline 
\end{tabular}
\label{table_obslog}
\\
$^{\rm a}$ Total exposure time of the observation (in $10^3$~s). \\
$^{\rm b}$ Number of individual exposures during the observation. \\
$^{\rm c}$ LWRS and MDRS are, respectively, large and medium 
\fuse\ apertures. \\  
\end{center}
\end{table*}

%\clearpage

\begin{table}
\begin{center}
\caption{Total interstellar column densities toward \hd9.}
\begin{tabular}{lc|lc}
\hline
\hline
species & $\log N$(\cmmd)\tablenotemark{a} & 
   species & $\log N$(\cmmd)\tablenotemark{a} \\
\hline
H$_2$ ($J=0$) & $19.61\pm0.04$  &  \ion{H}{1}   & $21.17\pm0.10$ \\
H$_2$ ($J=1$) & $19.62\pm0.03$  &  \ion{D}{1}   & $16.16\pm0.12$ \\
H$_2$ ($J=2$) & $18.15\pm0.25$  &  \ion{O}{1}   & $17.93\pm0.10$ \\
H$_2$ ($J=3$) & $17.32\pm0.30$  &  \ion{N}{1}   & $17.10\pm0.04$ \\
H$_2$ ($J=4$) & $15.44\pm0.08$  &  \ion{Fe}{2}  & $15.22\pm0.04$ \\
H$_2$ ($J=5$) & $14.81\pm0.06$  &  \ion{P}{2}   & $14.58\pm0.07$ \\
H$_2$ ($J=6$) & $13.60\pm0.10$  &  \ion{C}{1}   & $14.1$:\\
H$_2$ (total) & $19.92\pm0.04$  &  \ion{C}{1*}  & $13.7$:\\
CO ($J=0$)    & $13.49\pm0.08$  &  \ion{C}{1**} & $13.2$:\\
CO ($J=1$)    & $13.35\pm0.12$  &  \ion{Ar}{1}  & $15.4$: \\
CO (total)    & $13.73\pm0.08$  &  HD (total)   & $14.52\pm0.12$  \\
\hline
\end{tabular}
\label{table_columns}
\\
$^{\rm a}$ $2\sigma$ error~bars. Colons indicate uncertain results. \\
\end{center}
\end{table}

%\clearpage

\begin{table}
\begin{center}
\caption{Ratios toward \hd9.}
\begin{tabular}{lc}
\hline
\hline
Ratio & Value$^{\rm a}$ \\
\hline
D/O & $(1.7\pm0.7)\times10^{-2}$ \\
D/N & $(1.1\pm0.4)\times10^{-1}$ \\
D/H & $(9.8\pm3.8)\times10^{-6}$ \\
O/H & $(5.8\pm2.0)\times10^{-4}$ \\
N/H & $(8.5\pm2.2)\times10^{-5}$ \\
\hline
CO/H$_2$ & $(6.5\pm1.4)\times10^{-7}$ \\
$f($H$_2)$ & $10.1\pm2.6$~\% \\
HD/\ion{D}{1}    & $2.3\pm1.0$~\% \\
HD/2H$_2$ & $(2.0\pm0.6)\times10^{-6}$ \\
\hline
\end{tabular}
\label{table_ratios}
\\
$^{\rm a}$ $2\sigma$ error~bars.
\end{center}
\end{table}

\end{document}